\documentclass[10pt,aps,fleqn,superscriptaddress,notitlepage,
  nofootinbib,preprintnumbers,nobalancelastpage]{revtex4-1}
\pdfoutput=1
\usepackage{amsmath,amssymb,pstricks,graphicx,xspace,subfigure}

\newcommand{\Pythia}{P\protect\scalebox{0.8}{YTHIA}\xspace}
\newcommand{\Sherpa}{S\protect\scalebox{0.8}{HERPA}\xspace}

\newcommand{\Dire}{D\protect\scalebox{0.8}{IRE}\xspace}
\newcommand{\eps}{\varepsilon}
\newcommand{\mc}[1]{\mathcal{#1}}
\newcommand{\mr}[1]{\mathrm{#1}}

\usepackage{hyperref}
\hypersetup{
  pdfauthor={Stefan Hoeche, Stefan Prestel},
  pdftitle={Triple collinear emissions in parton showers},
  pdfkeywords={QCD, Parton Shower, NLO}
}
\begin{document}
\preprint{SLAC-PUB-16963}
\preprint{FERMILAB-PUB-17-122-T}
\preprint{MCNET-17-05}
\title{Triple collinear emissions in parton showers}
\author{Stefan~H{\"o}che}
\affiliation{SLAC National Accelerator Laboratory,
  Menlo Park, CA, 94025, USA}
\author{Stefan~Prestel}
\affiliation{Fermi National Accelerator Laboratory,
  Batavia, IL, 60510-0500, USA}
\begin{abstract}
  A framework to include triple collinear splitting functions into parton showers is
  presented, and the implementation of flavor-changing NLO splitting kernels is discussed
  as a first application. The correspondence between the Monte-Carlo integration and the
  analytic computation of NLO DGLAP evolution kernels is made explicit for both timelike
  and spacelike parton evolution. Numerical simulation results are obtained with two
  independent implementations of the new algorithm, using the two independent
  event generation frameworks \Pythia and \Sherpa.
\end{abstract}
\maketitle

\section{Introduction}
Parton showers solve the leading-order DGLAP equations~\cite{Gribov:1972ri,Lipatov:1974qm,
  Dokshitzer:1977sg,Altarelli:1977zs} using Markovian Monte-Carlo algorithms~\cite{Buckley:2011ms}.
As such they work at much lower computational precision than many other calculational tools
used in high-energy physics to date~\cite{Badger:2016bpw}. Due to their importance for both
experimental analyses and phenomenological surveys, a limited set of the most important higher-order
effects has been included into parton showers over time, such as angular ordering~\cite{Marchesini:1987cf},
and soft-gluon enhancement~\cite{Catani:1990rr}. The numerical size of the remaining
theoretical uncertainties is unclear, especially since parton showers are tuned to match
the most relevant experimental observables. The net effect of this tuning is that their
predictions are most often accurate, yet imprecise, and that the level of imprecision is
difficult to quantify numerically.
As fully exclusive, high precision simulations are mandatory in order to perform reliable
measurements of Standard Model parameters and/or searches for physics beyond the Standard Model,
the extension of parton showers to higher formal accuracy would benefit large parts
of the high-energy physics community.

The possibility of including next-to-leading order corrections into parton showers
has been explored early on~\cite{Kato:1986sg,Kato:1988ii,Kato:1990as,Kato:1991fs}
and was revisited recently~\cite{Hartgring:2013jma,Li:2016yez}. NLO splitting functions
have been recomputed using a novel regularization scheme~\cite{Jadach:2011kc,Gituliar:2014eba}.
The dependence of NLO matching terms on the parton-shower evolution scheme has been
investigated in detail~\cite{Jadach:2016zgk}. In addition, the first solutions to incorporate
effects beyond the leading-color approximation into parton showers have been
found~\cite{Platzer:2012hp,Nagy:2015hwa}, and threshold logarithms have been
included in a fully automated approach~\cite{Nagy:2016pwq}. 

In this publication, we construct a framework for the simulation of triple-collinear
parton splittings, which contribute to the next-to-leading order corrections to DGLAP
evolution~\cite{Curci:1980uw,Furmanski:1980cm,Floratos:1980hk,Floratos:1980hm}.
Triple-collinear splitting functions have been known since long~\cite{Catani:1999ss},
but they have not been included into parton showers to date\footnote{First ideas to include
  $2\to 4$ branchings in final-state evolution were presented in~\cite{Li:2016yez}.}.
We start with the simplest case of the flavor-changing splitting kernels. We use these $1\to 3$
kernels to recompute the timelike and spacelike NLO splitting functions $P_{qq'}$ in the
$\overline{\mr{MS}}$ scheme, and we show how the result can be implemented straightforwardly
in its differential form in a Markovian Monte-Carlo simulation, such that the integral
matches $P_{qq'}$ up to momentum conserving effects. Our algorithm depends crucially
on the usage of a weighted parton shower, a technique that was presented
in~\cite{Hoeche:2009xc,Lonnblad:2012hz}. We see an opportunity to extend our
new method to more complicated triple-collinear splitting functions, and
to include virtual corrections, such that all NLO kernels may eventually be
calculated on-the-fly, similar to the computation of a fixed-order result
in the dipole subtraction method~\cite{Catani:1996vz}.

The outline of this publication is as follows: Sec.~\ref{sec:ps_formalism} highlights
the correspondence between the formalisms for parton-shower evolution and DGLAP evolution.
The main components of parton-showers are the splitting kernels and the kinematics mapping,
which define the probability and kinematics in the transition from an $n$-parton final state
to an $n+1$-parton final state. Section~\ref{sec:formalism} therefore presents the recomputation
of the timelike and spacelike NLO splitting kernels $P_{qq'}$ and, based on the individual terms
identified in the analytical calculation, the construction of a formalism to include $1\to 3$
branchings in the parton shower. We present a validation of our numerical implementation and
a test of the numerical impact of $q\to q'$ and $q\to\bar{q}$ splittings in Sec.~\ref{sec:results}.
The kinematical mappings introduced to simulate $1\to 3$ splittings are an integral part of
the new algorithm, but their presentation is rather technical and has therefore been
included in App.~\ref{sec:phasespace}. Section~\ref{sec:conclusions} contains some
concluding remarks.

\section{Parton-shower formalism}
\label{sec:ps_formalism}
Parton showers implement QCD evolution equations, most commonly the DGLAP equation~\cite{
  Gribov:1972ri,Lipatov:1974qm,Dokshitzer:1977sg,Altarelli:1977zs}, which governs the evolution 
in the limits of collinear initial- and final-state parton branchings. The main components of
a parton-shower are thus the evolution or splitting kernels and the kinematics mapping
which defines how an $n$-parton final state transitions to an $n+1$-parton final state.
Modern parton showers implement local four-momentum conservation during this transition,
which requires the presence of a parton (or a set of partons) that compensate the missing
energy when the parton undergoing evolution is taken off its mass shell. Most commonly
this so-called recoil partner is identified with the color-connected parton in the large-$N_c$
limit. In order to construct a parton shower implementing triple-collinear splitting functions
we are thus left with two main tasks: One is to show that such a shower will implement the
NLO DGLAP evolution kernels that pertain to the triple-collinear parton branchings.
The other is to define kinematics mappings that allow us to generate $1\to 3$ transitions
in the presence of a recoil partner. We will address the problem of the connection of the
parton-shower formalism to the DGLAP equation in this section, while the definition of the
kinematics and a derivation of the related phase-space factorization in $D$ dimensions
is presented in App.~\ref{sec:phasespace}. We will make use of both results in
Sec.~\ref{sec:formalism}.

The evolution of parton densities and fragmentation functions in the collinear limit 
is governed by the DGLAP equations~\cite{Gribov:1972ri,Lipatov:1974qm,Dokshitzer:1977sg,Altarelli:1977zs}.
While they are schematically similar for initial and final state, the implementation in
parton-shower programs is radically different between the two, owing to the fact that
Monte-Carlo simulations are typically performed for inclusive final states. Nevertheless
parton showers do solve the DGLAP equations both in timelike and in spacelike evolution.
We will start with the evolution equations for the fragmentation functions $D_a^{h}(x,Q^2)$ for
parton of type $a$ to fragment into hadron $h$, and we will suppress the index $h$ for brevity,
\begin{equation}\label{eq:pdf_evolution}
  \frac{{\rm d}\,xD_{a}(x,t)}{{\rm d}\ln t}=
  \sum_{b=q,g}\int_0^1{\rm d}\tau\int_0^1{\rm d} z\,\frac{\alpha_s}{2\pi}
  \big[zP_{ab}(z)\big]_+\,\tau D_{b}(\tau,t)\,\delta(x-\tau z)\;.
\end{equation}
In this context, $P_{ab}$ are the unregularized DGLAP evolution kernels, which can be expanded
into a power series in the strong coupling. The plus prescription can be used to enforce the 
momentum and flavor sum rules:
\begin{equation}\label{eq:sf_regularization}
  \big[zP_{ab}(z)\big]_+=\lim\limits_{\eps\to 0}zP_{ab}(z,\eps)\;,
\end{equation}
where
\begin{equation}
    P_{ab}(z,\eps)=P_{ab}(z)\,\Theta(1-z-\eps)
    -\delta_{ab}\sum_{c\in\{q,g\}}
    \frac{\Theta(z-1+\eps)}{\eps}
    \int_0^{1-\eps}{\rm d}\zeta\,\zeta\,P_{ac}(\zeta)\;.
\end{equation}
For finite $\eps$, the endpoint subtraction in Eq.~\eqref{eq:sf_regularization} 
can be interpreted as the approximate virtual plus unresolved real corrections, 
which are included in the parton shower because the Monte-Carlo algorithm
naturally implements a unitarity constraint~\cite{Jadach:2003bu}. The precise value 
of $\eps$ in this case is defined in terms of an infrared cutoff on the 
evolution variable, using four-momentum conservation. For $0<\eps\ll 1$,
Eq.~\eqref{eq:pdf_evolution} changes to
\begin{equation}\label{eq:pdf_evolution_constrained}
  \frac{1}{D_{a}(x,t)}\,\frac{{\rm d} D_{a}(x,t)}{{\rm d}\ln t}=
  -\sum_{c=q,g}\int_0^{1-\eps}{\rm d}\zeta\,\zeta\,\frac{\alpha_s}{2\pi}P_{ac}(\zeta)\,
  +\sum_{b=q,g}\int_x^{1-\eps}\frac{{\rm d} z}{z}\,
  \frac{\alpha_s}{2\pi}\,P_{ab}(z)\,\frac{D_{b}(x/z,t)}{D_{a}(x,t)}\;.
\end{equation}
Using the Sudakov form factor
\begin{equation}\label{eq:sudakov}
  \Delta_a(t_0,t)=\exp\bigg\{-\int_{t_0}^{t}\frac{{\rm d} \bar{t}}{\bar{t}}
  \sum_{c=q,g} \int_0^{1-\eps}{\rm d}\zeta\,\zeta\,\frac{\alpha_s}{2\pi}P_{ac}(\zeta)\bigg\}
\end{equation}
one can define the generating function for splittings of parton $a$ as
\begin{equation}\label{eq:def_updf}
  \mc{D}_a(x,t,\mu^2)=D_a(x,t)\Delta_a(t,\mu^2)\;.
\end{equation}
Equation~\eqref{eq:pdf_evolution_constrained} can now be written in the simple form
\begin{equation}\label{eq:pdf_evolution_constrained_2}
  \frac{{\rm d}\ln\mc{D}_a(x,t,\mu^2)}{{\rm d}\ln t}
  =\sum_{b=q,g}\int_x^{1-\eps}\frac{{\rm d} z}{z}\,
  \frac{\alpha_s}{2\pi}\,P_{ab}(z)\,\frac{D_{b}(x/z,t)}{D_{a}(x,t)}\;.
\end{equation}
The generalization to an $n$-parton state, $\vec{a}=\{a_1,\ldots,a_n\}$,
with jets and incoming hadrons resolved at scale $t$ can be made in terms
of PDFs, $f$, and fragmenting jet functions, $\mc{G}$~\cite{Procura:2009vm,Jain:2011xz}.
We define the generating function for this state as
$\mc{F}_{\vec{a}}(\vec{x},t,\mu^2)$. It obeys the evolution equation
\begin{equation}\label{eq:pdf_evolution_constrained_3}
  \begin{split}
  \frac{{\rm d}\ln\mc{F}_{\vec{a}}(\vec{x},t,\mu^2)}{{\rm d}\ln t}
  =&\sum_{i\in{\rm IS}}\sum_{b=q,g}\int_{x_i}^{1-\eps}\frac{{\rm d} z}{z}\,
  \frac{\alpha_s}{2\pi}\,P_{ba_i}(z)\,\frac{f_{b}(x_i/z,t)}{f_{a_i}(x_i,t)}\\
  &\qquad+\sum_{j\in{\rm FS}}\sum_{b=q,g}\int_{x_j}^{1-\eps}\frac{{\rm d} z}{z}\,
  \frac{\alpha_s}{2\pi}\,P_{a_jb}(z)\,\frac{\mc{G}_{b}(x_j/z,t)}{\mc{G}_{a_j}(x_j,t)}\;.
  \end{split}
\end{equation}
This equation can be solved using Markovian Monte-Carlo techniques known as parton showers~\cite{Buckley:2011ms}.
In most cases, parton showers implement final-state branchings in unconstrained evolution.
Since Eq.~\eqref{eq:pdf_evolution_constrained} also applies to $\mc{G}$~\cite{Procura:2009vm}, we can use 
Eq.~\eqref{eq:pdf_evolution_constrained} to remove the dependence of $\mc{G}$ from 
Eq.~\eqref{eq:pdf_evolution_constrained_3}, thus leading to the differential
branching probability
\begin{equation}\label{eq:pdf_evolution_constrained_4}
  \begin{split}
    \frac{{\rm d}}{{\rm d}\ln t}\ln\bigg(
    \frac{\mc{F}_{\vec{a}}(\vec{x},t,\mu^2)}{\prod_{j\in\rm FS}\mc{G}_{a_j}(x_j,t)}\bigg)
  =&\sum_{i\in{\rm IS}}\sum_{b=q,g}\int_{x_i}^{1-\eps}\frac{{\rm d} z}{z}\,
  \frac{\alpha_s}{2\pi}\,P_{ba_i}(z)\,\frac{f_{b}(x_i/z,t)}{f_{a_i}(x_i,t)}
  +\sum_{j\in{\rm FS}}\sum_{b=q,g}\int_0^{1-\eps}{\rm d} z\,z\,
  \frac{\alpha_s}{2\pi}\,P_{a_jb}(z)\;.
  \end{split}
\end{equation}
A direct consequence of this relation is that the Sudakov factor, Eq.~\eqref{eq:sudakov},
must be used in final-state parton showers that implement splitting kernels beyond the
leading order, or else the sum rules will be violated~\cite{Jadach:2003bu}. 
However, at leading order the additional factor $\zeta$ in the integral of
Eq.~\eqref{eq:sudakov} can be replaced by a symmetry factor, because the
leading-order DGLAP splitting functions, $P_{ab}^{(0)}$, obey the symmetries
\begin{equation}\label{eq:lokernels_symmetry}
  \begin{split}
    \sum_{b=q,g}\int_{0}^{1-\eps} {\rm d}z\, z\, P_{qb}^{(0)}(z)
    =&\int_{\eps}^{1-\eps} {\rm d}z\, P_{qq}^{(0)}(z)+\mc{O}(\eps)\;,\\
    \sum_{b=q,g}\int_{0}^{1-\eps} {\rm d}z\, z\, P_{gb}^{(0)}(z)
    =&\int_{\eps}^{1-\eps} {\rm d}z\,\Big[\;
      \frac{1}{2}P_{gg}^{(0)}(z)+n_f\,P_{gq}^{(0)}(z)\;\Big]+\mc{O}(\eps)\;.
  \end{split}
\end{equation}
This relates the branching formalism employed for our new parton shower to the
conventional technique for final-state parton evolution~\cite{Buckley:2011ms},
where the factor $\zeta$ is replaced by $1/2$. The new formalism has a convenient
physical interpretation: The factor $\zeta$ identifies the final-state parton
undergoing evolution in the same way that the initial-state parton is identified
in initial-state evolution.
We will make use of this result in Sec.~\ref{sec:sub_impl}, where we show
how to implement the differential form of the integrated splitting kernels
computed in Sec.~\ref{sec:pqq_analytic}.

\section{Incorporation of 1$\to$3 branchings}
\label{sec:formalism}
In this section we detail the formalism used to implement triple-collinear
splitting functions, both in the spacelike and in the timelike case. The main result
is given by Eq.~\eqref{eq:tcps_mc}, which unsurprisingly bears a remarkable similarity
to the formulation of a fixed-order NLO calculation in the subtraction method.
Our algorithm must satisfy the constraint that the integral over the splitting
function evaluates to the corresponding NLO evolution kernel first computed
in~\cite{Curci:1980uw,Furmanski:1980cm,Floratos:1980hk,Floratos:1980hm}
and rederived in~\cite{Heinrich:1997kv,Ellis:1996nn}.
To verify this, we recompute the flavor-changing timelike and spacelike kernels
$P_{qq'}$ in Sec.~\ref{sec:pqq_analytic}. We then identify the relevant
components to be implemented in the Monte-Carlo simulation and comment on
the appropriate transformation of the MC integration variables listed in
App.~\ref{sec:phasespace}. We also comment on the possibility to extend this method
to splitting functions with leading-order contributions and virtual corrections.

In the triple collinear limit of partons $a$, $i$ and $j$, any QCD (associated)
matrix element squared factorizes as~\cite{Catani:1999ss}
\begin{equation}\label{eq:tc_me_factorization}
  |M_{a,i,j,\ldots,k,\ldots}(p_a,p_i,p_j,\ldots)|^2=
  \left(\frac{8\pi\mu^{2\eps}\alpha_s}{s_{aij}}\right)^2
  \mc{T}^{ss'}_{aij,\ldots}(p_{aij},\ldots)\,
  P^{ss'}_{aij}(p_a,p_i,p_j)\;,
\end{equation}
where the superscripts denote the spin-dependence of both the splitting function
and the reduced matrix element. We will implement the spin-averaged splitting functions,
$\langle P_{aij}\rangle(p_a,p_i,p_j)$, together with related counterterms that are
identified in Secs.~\ref{sec:local_sub} and~\ref{sec:sub_impl}. The factor in
parentheses in Eq.~\eqref{eq:tc_me_factorization} is common to all terms. The two powers
of the strong coupling are both evaluated at the parton-shower evolution variable, $t$.
One factor $s_{aij}$ will be combined with the last term in Eqs.~\eqref{eq:tc_ff_ps},
\eqref{eq:tc_fi_ps},~\eqref{eq:tc_if_ps} and~\eqref{eq:tc_ii_ps}, while the other cancels
after transformation of the $s_{ai}$ integration using Eqs.~\eqref{eq:sai_v_ff} 
and~\eqref{eq:sai_v_if}. We will comment on this in Sec.~\ref{sec:sub_impl}.

\subsection{Fixed-order calculation}
\label{sec:pqq_analytic}
We use the method outlined in \cite{Ritzmann:2014mka} to compute both the timelike 
and the spacelike flavor-changing NLO splitting kernels for massless partons
\begin{equation}\label{eq:p1_qqp}
  \begin{split}
    P_{qq'}^{(T)}(z)=&\;C_F T_R\left(+(1+z)\log^2(z)-\left(\frac{8}{3}z^2+9z+5\right)\log(z)+\frac{56}{9}z^2+4z-8-\frac{20}{9z}\right)\;,\\
    P_{qq'}^{(S)}(z)=&\;C_F T_R\left(-(1+z)\log^2(z)-\left(\frac{8}{3}z^2+5z+1\right)\log(z)-\frac{56}{9}z^2+6z-2+\frac{20}{9z}\right)\;.
  \end{split}
\end{equation}
The timelike splitting functions can be extracted from the term proportional to $\delta(s)$
in the two-loop matching of the fragmenting jet function, $\mc{G}$, while the spacelike 
splitting function is obtained similarly from the $\delta(s)$ term in the matching condition
of the beam function. In the timelike case, the matching condition reads
\begin{equation}\label{eq:twoloop_ffmatch}
  \begin{split}
    \mc{G}_q^{i(2)}(s,z,\mu)=\mc{J}_{qi}^{(2)}(s,z,\mu)
    +\sum_j\int_z^1\frac{{\rm d}x}{x}\mc{J}_{qj}^{(1)}(s,z/x,\mu)D_j^{i(1)}(x,\mu)
    +\delta(s)D_q^{i(2)}(z,\mu)\;.
  \end{split}
\end{equation}
The perturbative fragmentation function at $\mc{O}(\alpha_s^2)$ is given by
\begin{equation}\label{eq:twoloop_ppdf}
  \begin{split}
    &D_j^{i(2)}(z,\mu)=\delta_{ij}\delta(1-z)
    -\frac{\alpha_s}{2\pi}\frac{1}{\eps}P_{ji}^{(0)}(z)\\
    &\qquad+\left(\frac{\alpha_s}{2\pi}\right)^2\left[
    -\frac{1}{2\eps}P_{ji}^{(1)}(z)+\frac{\beta_0}{4\eps^2}P_{ji}^{(0)}(z)
    +\frac{1}{2\eps^2}\int_z^1\frac{{\rm d}x}{x}
    P_{jk}^{(0)}(x)P_{ki}^{(0)}(z/x)\right]\;.
  \end{split}
\end{equation}
In the timelike case we employ the phase-space parametrization 
of~\cite{Gehrmann-DeRidder:2003pne}. We factor out the two-particle
phase space, the integration over the three-particle invariant 
$y_{aij}=s_{aij}/q^2$ and the corresponding factors $(y_{aij}(1-y_{aij}))^{1-2\eps}$
as well as the integration over one of the light-cone momentum fractions,
which is chosen to be $\tilde{z}=s_{ak}/q^2/(1-y_{aij})$. We also remove the square
of the normalization factor $(4\pi)^{\eps}/(16\pi^2\Gamma(1-\eps))\,(q^2)^{1-\eps}$.
The remaining one-emission phase-space integral reads
\begin{equation}\label{eq:tcps_tl}
  \begin{split}
    \int{\rm d}\Phi_{+1}^{(F)}=&\;
    (1-\tilde{z})^{1-2\eps}\tilde{z}^{-\eps}
    \int_0^1{\rm d}\tau\,(\tau(1-\tau))^{-\eps}
    \int_0^1{\rm d}v\,(v(1-v))^{-\eps}\;
    \frac{\Omega(1-2\eps)}{\Omega(2-2\eps)}
    \int_0^1{\rm d}\chi\,2(4\chi(1-\chi))^{-1/2-\eps}\;,
  \end{split}
\end{equation}
where $\Omega(n)=2\pi^{n/2}/\Gamma(n/2)$.
The variables $\tau$ and $v$ are given by the transformation\footnote{%
  Note that we define $\tilde{z}_j\to(1-\tilde{z})\tau$, while the corresponding
  transformation in~\cite{Gehrmann-DeRidder:2003pne} reads
  $\tilde{z}\to(1-\tilde{z}_j)\tau$.}
\begin{equation}
  s_{ai}=s_{aij}(1-\tilde{z}_j)\,v\;,
  \qquad
  \tilde{z}_j=\frac{s_{jk}/q^2}{1-y_{aij}}=(1-\tilde{z})\,\tau\;.
\end{equation}
The azimuthal angle integration is parametrized using $\chi$, which is defined as
$ 
  s_{ij}=s_{ij,-}+\chi(s_{ij,+}-s_{ij,-})\,,
$ 
with $s_{ij,\pm}$ being the two solutions of the quadratic equation
$\cos^2\phi_{a,i}^{j,k}=1$, cf.~Eq.~\eqref{eq:tc_ff_ps_dd_cosphi}%
~\cite{Gehrmann-DeRidder:2003pne}.

We can now integrate the only diagram contributing to the timelike NLO DGLAP
kernel, $P_{qq'}^{(T)}(\tilde{z})$, which is given by the triple collinear
splitting function~\cite{Catani:1999ss}
\begin{equation}\label{eq:pqqp_tc}
  P_{qq'}^{1\to3}=\frac{1}{2}C_FT_R\frac{s_{aij}}{s_{ai}}
  \left[-\frac{t_{ai,j}^2}{s_{ai}s_{aij}}+\frac{4\,\tilde{z}_j+(\tilde{z}_a-\tilde{z}_i)^2}{\tilde{z}_a+\tilde{z}_i}
    +(1-2\varepsilon)\left(\tilde{z}_a+\tilde{z}_i-\frac{s_{ai}}{s_{aij}}\right)\right]\;,
\end{equation}
where $(\tilde{z}_a+\tilde{z}_i)\,t_{ai,j}=2(\tilde{z}_as_{ij}-\tilde{z}_is_{aj})+(\tilde{z}_a-\tilde{z}_i)s_{ai}$.
The result is
\begin{equation}\label{eq:pqqp_int_tl}
  \begin{split}
    &\frac{1}{C_FT_R}\int{\rm d}\Phi_{+1}^{(F)}P_{qq'}^{1\to3}=
    -\frac{1}{\eps}\left(2(1+\tilde{z})\log\tilde{z}+(1-\tilde{z})+\frac{4}{3\tilde{z}}(1-\tilde{z}^3)\right)\\
    &\qquad-4(1+\tilde{z})\Big({\rm Li}_2(\tilde{z})-\zeta_2\Big)+3(1+\tilde{z})\log^2\tilde{z}
    -\frac{16}{3}(1-\tilde{z})+\frac{2}{3\tilde{z}}(1-\tilde{z}^3)\\
    &\qquad+\left(\frac{8}{3\tilde{z}}+\tilde{z}+3\right)\log\tilde{z}
    +\left(\frac{8}{3\tilde{z}}(1-\tilde{z}^3)+2(1-\tilde{z})\right)\log(1-\tilde{z})
    +\mc{O}(\eps)\;.
  \end{split}
\end{equation}
Upon including the propagator term from Eq.~\eqref{eq:tc_me_factorization} and
the phase-space factor $y_{aij}^{1-2\eps}$, the leading pole will receive an additional
factor $-\delta(y_{aij})/2\eps$. The $1/\eps^2$ coefficient thus generated is removed
by the renormalization of the fragmentation function. As required, it agrees up to a sign
with the corresponding second order $1/\eps^2$ coefficient in Eq.~\eqref{eq:twoloop_ppdf},
which we write as
\begin{equation}\label{eq:ffpdf_ren}
  \mc{P}_{qq'}(\tilde{z})=\int_{\tilde{z}}^1\frac{{\rm d}x}{x}
  P_{qg}^{(0)}(x)P_{gq}^{(0)}(z/x)=\,C_F T_R\left(2(1+\tilde{z})\log\tilde{z}
  +(1-\tilde{z})+\frac{4}{3\tilde{z}}(1-\tilde{z}^3)\right)\;.
\end{equation}
Equation~\eqref{eq:twoloop_ffmatch} can now be employed to extract the NLO DGLAP kernel
$P_{qq'}^{(T)}(z)$ from the finite remainder of Eq.~\eqref{eq:pqqp_int_tl}. We subtract
the corresponding convolution of the one-loop matching coefficient with the one-loop
fragmentation function, which is given by
\begin{equation}\label{eq:match_tl}
  \mc{I}_{qq'}^{(F)}(\tilde{z})=2\int_{\tilde{z}}^1\frac{{\rm d}x}{x}
  C_F\left(\frac{1+(1-x)^2}{x}\log(x(1-x))+x\right)P_{gq}^{(0)}(\tilde{z}/x)\;.
\end{equation}
Using this technique, we finally obtain the result in Eq.~\eqref{eq:p1_qqp}.

We now proceed to perform the integral over the triple-collinear splitting function
in the spacelike case. We use the phase space parametrization in~\cite{Ellis:1996nn}.
The azimuthal angle integral is most conveniently parametrized using
Eq.~\eqref{eq:tc_ff_ps_dd_costh}, which gives
$ 
  {\rm d}\phi_{i,j}^{a,b}={\rm d}(\vec{p}_{i\perp}\vec{p}_{j\perp})/
    (|\vec{p}_{i\perp}||\vec{p}_{j\perp}|\sin\phi_{i,j}^{a,b})\,,
$ 
with $\vec{p}_{\perp}$ the transverse momenta with respect to the (anti-)collinear
directions defined by $p_a$ (and $p_b$). We can use a transformation identical to
the timelike case~\cite{Gehrmann-DeRidder:2003pne}. We define
$s_{ij}=s_{ij,-}+\chi(s_{ij,+}-s_{ij,-})$, where $s_{ij,\pm}$ are the two solutions
of the quadratic equation $\cos^2\phi_{i,j}^{a,b}=1$. The related angular integral is
$ 
  {\rm d}\phi_{i,j}^{a,b}(\sin^2\phi_{i,j}^{a,b})^{-\eps}=
  2{\rm d}\chi\,(4\chi(1-\chi))^{-1/2-\eps}\;.
$ 
We remove the normalization factor $(4\pi)^{2\eps}/(16\pi^2\Gamma(1-\eps))^2\,s_{aij}^{1-2\eps}$.
The full phase space relevant to our computation is then given by
\begin{equation}\label{eq:tcps_sl}
  \int{\rm d}\Phi_{+1}^{(I)}=\tilde{z}^{-1+\eps}
  \int_0^1{\rm d}x\,(1-x)^{-\eps}(x-\tilde{z})^{-\eps}
  \int_0^1{\rm d}v\,(v(1-v))^{-\eps}\,
  \frac{\Omega(1-2\eps)}{\Omega(2-2\eps)}
  \int_0^1{\rm d}\chi\,2(4\chi(1-\chi))^{-1/2-\eps}\;.
\end{equation}
Using Eq.~\eqref{eq:pqqp_tc} and the crossing relation
\begin{equation}\label{eq:tcsf_qqp}
  P_{qq'}(\tilde{z}_1,\tilde{z}_2,\tilde{z}_3,s_{12},s_{13},s_{23})=
  \tilde{z}_1\,P_{q'q}(1/\tilde{z}_1,-\tilde{z}_2/\tilde{z}_1,-\tilde{z}_3/\tilde{z}_1,
  -s_{12},-s_{13},s_{23})\;,
\end{equation}
we can integrate the only contribution to the spacelike NLO DGLAP
kernel $P_{qq'}^{(S)}(z)$. The result can be expressed in terms of
Eq.~\eqref{eq:pqqp_int_tl} (see also~\cite{Curci:1980uw,Dokshitzer:2005bf})
\begin{equation}\label{eq:pqqp_int_sl}
  \begin{split}
    \int{\rm d}\Phi_{+1}^{(I)}\,\tilde{z}P_{q'q}^{1\to3}=
    \int{\rm d}\Phi_{+1}^{(F)}P_{qq'}^{1\to3}
    -2\log\tilde{z}\int_{\tilde{z}}^1\frac{{\rm d}x}{x}
    P_{qg}^{(0)}(x)P_{gq}^{(0)}(\tilde{z}/x)+\mc{O}(\eps)\;.
  \end{split}
\end{equation}
As in the timelike case, the $1/\eps$ coefficient will eventually be removed
by the renormalization of the PDF. It agrees with the corresponding second order
$1/\eps^2$ coefficient $\mc{P}_{qq'}(\tilde{z})$ of Eq.~\eqref{eq:ffpdf_ren} and
with the corresponding coefficient in the timelike calculation. The finite remainder 
of Eq.~\eqref{eq:pqqp_int_sl} can be employed to extract the NLO DGLAP kernel
$P_{qq'}^{(S)}(z)$. In order to do so, we must subtract the corresponding convolution
of the one-loop matching coefficient with the first-order renormalization term
of the PDFs, which is given by
\begin{equation}\label{eq:match_sl}
  \mc{I}_{qq'}^{(I)}(\tilde{z})=2\int_{\tilde{z}}^1\frac{{\rm d}x}{x}
  T_R\Big((1-2x(1-x))\log(1-x)+2x(1-x)\Big)P_{qg}^{(0)}(\tilde{z}/x)\;.
\end{equation}
Using this technique, we finally obtain the result in Eq.~\eqref{eq:p1_qqp}.

The above computations allow us to obtain the NLO DGLAP splitting functions
using the triple-collinear splitting functions as an input. The drawback of this
method is that the calculation must be performed in $D=4-2\eps$ dimensions,
and that the cancellation of the singularities occurs between the integrals.
In the next section we will therefore construct a local subtraction scheme
that allows to cancel singularities at the integrand level and implement the
computation in a manner similar to standard subtraction~\cite{Catani:1996vz},
more precisely {\it modified} subtraction~\cite{Frixione:2002ik}.

\subsection{Definition of a local subtraction procedure}
\label{sec:local_sub}
We will now proceed to define a scheme for the fully numerical computation
of the kernels in Eq.~\eqref{eq:p1_qqp}. This method allows us to evaluate
the integrals leading to Eq.~\eqref{eq:p1_qqp} in four dimensions, which in
turn allows to use standard Monte-Carlo techniques to evaluate them numerically.
Our method can be likened to a standard NLO calculation using modified
subtraction techniques~\cite{Frixione:2002ik}. In this context, it is crucial
that divergences of the triple-collinear splitting functions cancel locally
against the subtraction terms. We therefore compute the differential radiation
pattern using the triple-collinear splitting functions of~\cite{Catani:1999ss},
subtracted by the spin-correlated iterated leading-order splitting functions
of~\cite{Somogyi:2005xz}. We then add the finite remainder of the integrated
leading-order splitting functions and the renormalization and matching
terms as an endpoint contribution. The details of this procedure are described
in the following.

Using the phase-space parametrizations in Eqs.~\eqref{eq:tcps_tl} and~\eqref{eq:tcps_sl}
we can compute the integrals of the iterated leading-order splitting kernels
corresponding to Eq.~\eqref{eq:tcsf_qqp}. This approximate kernel reads
\begin{equation}\label{eq:tcsf_qqpa}
  \begin{split}
  \tilde{P}_{qq'}^{1\to3}(\tilde{z}_a,\tilde{z}_i,\tilde{z}_j,s_{ai},s_{aj},s_{ij})=&\;
  \frac{s_{aij}}{s_{ai}}\,P_{qg}^{(0)}(\tilde{z}_j)\,P_{gq}^{(0)}\Big(\frac{\tilde{z}_a}{\tilde{z}_a+\tilde{z}_i}\Big)\\
  =&\;  C_F T_R\frac{s_{aij}}{s_{ai}}\left(\frac{1+\tilde{z}_j^2}{1-\tilde{z}_j}-\eps(1-\tilde{z}_j)\right)
  \left(1-\frac{2}{1-\eps}\frac{\tilde{z}_a\tilde{z}_i}{(\tilde{z}_a+\tilde{z}_i)^2}\right)\;.
  \end{split}
\end{equation}
Its integrals are given by
\begin{equation}\label{eq:pqqpa_int}
  \begin{split}
    \mr{I}_{qq'}^{(F)}(\tilde{z})=&\;\int{\rm d}\Phi_{+1}^{(F)}\tilde{P}_{qq'}^{1\to3}(\tilde{z},\Phi_{+1})\;\,=
    \int{\rm d}\Phi_{+1}^{(F)}P_{qq'}^{1\to3}(\tilde{z},\Phi_{+1})-\Delta\mr{I}_{qq'}(\tilde{z})+\mc{O}(\eps)\;,\\
    \mr{I}_{qq'}^{(I)}(\tilde{z})=&\;\int{\rm d}\Phi_{+1}^{(I)}\,\tilde{z}\tilde{P}_{q'q}^{1\to3}(1/\tilde{z},\Phi_{+1})=
    \int{\rm d}\Phi_{+1}^{(I)}\,\tilde{z}P_{q'q}^{1\to3}(1/\tilde{z},\Phi_{+1})-\Delta\mr{I}_{qq'}(\tilde{z})+\mc{O}(\eps)\;,\\
  \end{split}
\end{equation}
where
\begin{equation}
  \Delta\mr{I}_{qq'}(\tilde{z})=C_F T_R\Big(5(1-\tilde{z})+2(1+\tilde{z})\log\tilde{z}\Big)\;.
\end{equation}
As required, the $1/\eps$ poles agree with the integrals of the triple-collinear splitting
function, Eqs.~\eqref{eq:pqqp_int_tl} and~\eqref{eq:pqqp_int_sl}. The difference in the
finite part is identical in the timelike and the spacelike case. This suggest that
the approximate kernel, Eq.~\eqref{eq:tcsf_qqpa} can be used as a subtraction term
for the full triple-collinear splitting kernel, Eq.~\eqref{eq:tcsf_qqp}.
It is not, however, a {\it local} subtraction term, as the $1/\eps$ pole generated
by the $v$-integral cancels only after azimuthal integration. In order to construct
a local subtraction term, we employ the spin-dependent splitting function, $P_{qg}^{\mu\nu}$,
computed in~\cite{Somogyi:2005xz}, together with the standard spin-dependent LO splitting function,
$P_{gq}^{\mu\nu}$ 
\begin{equation}\label{eq:sd_lo_kernels}
  \begin{split}
    P_{qg}^{\mu\nu}(z,k_\perp)=&\;C_F\left[-\frac{2\,z}{1-z}\frac{k_\perp^\mu k_\perp^\nu}{k_\perp^2}
      +\frac{1-z}{2}\left(-g^{\mu\nu}+\frac{p^\mu n^\nu+p^\nu n^\mu}{pn}\right)\right]\;,\\
    P_{gq}^{\mu\nu}(z,k_\perp)=&\;T_R\left[-g^{\mu\nu}+4z(1-z)\frac{k_\perp^\mu k_\perp^\nu}{k_\perp^2}\right]\;.
  \end{split}
\end{equation}
Their scalar product generates an additional contribution to Eq.~\eqref{eq:tcsf_qqpa}, which reads
\begin{equation}\label{eq:pqqp_sd_ct}
  \Delta\tilde{P}_{qq'}^{1\to3}(\tilde{z}_a,\tilde{z}_i,\tilde{z}_j,s_{ai},s_{aj},s_{ij})=
  C_FT_R \frac{s_{aij}}{s_{ai}}\frac{4 \tilde{z}_a \tilde{z}_i \tilde{z}_j}{(1-\tilde{z}_j)^3}\left(1-2\cos^2\phi_{a,j}^{i,k}\right)\;.
\end{equation}
The modified approximate kernel exactly cancels the $1/s_{ai}$ poles present in the
triple collinear splitting function, such that their difference can be integrated
in four dimensions, leading to the expected result
\begin{equation}\label{eq:spqqp_int}
  \begin{split}
    \int{\rm d}\Phi_{+1}^{(F)}\left(P_{qq'}^{1\to3}
    -\tilde{P}_{qq'}^{1\to3}-\Delta\tilde{P}_{qq'}^{1\to3}\right)(\tilde{z},\Phi_{+1})=&\;
    \Delta\mr{I}_{qq'}(\tilde{z})\;,\\
    \int{\rm d}\Phi_{+1}^{(I)}\,\tilde{z}\left(P_{qq'}^{1\to3}
    -\tilde{P}_{qq'}^{1\to3}-\Delta\tilde{P}_{qq'}^{1\to3}\right)(\tilde{z},\Phi_{+1})=&\;
    \Delta\mr{I}_{qq'}(\tilde{z})\;.
  \end{split}
\end{equation}
We now define the functions
\begin{equation}\label{eq:tcps_def_rs}
  \begin{split}
    \mr{R}_{qq'}^{(F)}(\tilde{z},\Phi_{+1})=&\;P_{qq'}^{1\to3}(\tilde{z},\Phi_{+1})\;,
    &\mr{S}_{qq'}^{(F)}(\tilde{z},\Phi_{+1})=&\;\tilde{P}_{qq'}^{1\to3}(\tilde{z},\Phi_{+1})
    +\Delta\tilde{P}_{qq'}^{1\to3}(\tilde{z},\Phi_{+1})\;,\\
    \mr{R}_{qq'}^{(I)}(\tilde{z},\Phi_{+1})=&\;\tilde{z}P_{q'q}^{1\to3}(1/\tilde{z},\Phi_{+1})\;,
    &\mr{S}_{qq'}^{(I)}(\tilde{z},\Phi_{+1})=&\;\tilde{z}\tilde{P}_{qq'}^{1\to3}(1/\tilde{z},\Phi_{+1})
    +\tilde{z}\Delta\tilde{P}_{qq'}^{1\to3}(1/\tilde{z},\Phi_{+1})\;.
  \end{split}
\end{equation}
This allows us to write the NLO kernel as
\begin{equation}\label{eq:tcps_mc}
  P_{qq'}^{(T/S)}(\tilde{z})=\Big(\mr{I}+\frac{1}{\eps}\,\mc{P}-\mc{I}\Big)_{qq'}^{(F/I)}(\tilde{z})+
  \int{\rm d}\Phi_{+1}^{(F/I)}(\mr{R}-\mr{S})_{qq'}^{(F/I)}(\tilde{z},\Phi_{+1})\;.
\end{equation}
This equation bears similarity to the definition of standard and hard events in the MC@NLO
method~\cite{Frixione:2002ik} without the related shower evolution. However, in our case it is
implemented not as a matching coefficient, but in the {\it exponent} of the all-orders Sudakov
form factor.

In fact, the parton-shower that is added explicitly in MC@NLO is already present in our case,
as we also include the leading-order simulation, which schematically generates the additional
contributions at $\mc{O}(\alpha_s^2)$
\begin{equation}\label{eq:mcnlo_ct}
  \begin{split}
    &\int{\rm d}\Phi_{+1}^{(F)}\tilde{\mc{P}}_{qq'}^{1\to 3}(\tilde{z},\Phi_{+1})
    \Big(O(\tilde{z},\Phi_{+1})-O(\tilde{z})\Big)\;,\\
    &\int{\rm d}\Phi_{+1}^{(I)}\,\tilde{z}\tilde{\mc{P}}_{qq'}^{1\to 3}(1/\tilde{z},\Phi_{+1})
    \Big(O(\tilde{z},\Phi_{+1})-O(\tilde{z})\Big)\;.
  \end{split}
\end{equation}
In this equation, $O$ stands for an arbitrary observable, which picks up the real-emission
phase-space dependence in the emission term of the parton shower, and the Born phase-space
dependence in the corresponding approximate virtual correction implemented through the
Sudakov form factor. As in the case of MC@NLO, Eq.~\eqref{eq:mcnlo_ct} provides the necessary
counterterms to generate the correct observable dependence on the real-emission phase-space
in Eq.~\eqref{eq:tcps_mc}. This allows to generate events which are distributed according
to the fully differential radiation pattern, as given by the triple-collinear splitting function.

In this context it is important to note that our leading-order parton shower 
does not yet include the spin-correlation term given by Eq.~\eqref{eq:pqqp_sd_ct}.
Therefore, the cancellation generated between terms from Eq.~\eqref{eq:mcnlo_ct}
and Eq.~\eqref{eq:tcps_mc} is non-local in the azimuthal angle. However, this effect
will be suppressed in practice, due to the fact that Eq.~\eqref{eq:mcnlo_ct} is large
only in the soft region $z_j\to1$, which is most often not resolved in experimental and
phenomenological analyses. We will address the implementation of Eq.~\eqref{eq:pqqp_sd_ct}
in the leading-order simulation in a future publication.

The form of Eq.~\eqref{eq:tcps_mc} suggests that our method generalizes to the case with
Born contribution and virtual corrections, and that the generic structure will be that of a
computation using the NLO dipole subtraction method~\cite{Catani:1996vz}, except that
the subtraction terms are evaluated in the real-emission phase space, as required for
generating parton-shower input configurations in an MC@NLO~\cite{Frixione:2002ik}.
A complete set of local counterterms for the real-emission contribution could then be
obtained from~\cite{Somogyi:2005xz}\footnote{We note that in the general case 
  the implementation will depend on the parton-shower evolution variable, as
  the phase-space factors in Eqs.~\eqref{eq:dcdc_ps_fs} and~\eqref{eq:dcdc_ps_is}
  will contribute additional logarithmic terms when expanded to $\mc{O}(\eps)$
  and combined with the leading pole arising from the soft gluon singularity.
  In addition, the functions $\mc{P}$ and $\mc{I}$ are renormalization scheme
  dependent. A change of renormalization scheme can be accommodated by redefining
  these terms.}.

\subsection{Implementation in the parton shower}
\label{sec:sub_impl}
This section describes the implementation of the local subtraction procedure outlined
above into a Monte-Carlo event generator. As opposed to a leading-order simulation,
where all splittings have $2\to 3$ kinematics, the new simulation includes an
integral over $2\to 4$ configurations, and endpoint contributions arising from
$(\mr{I}+\mc{P}/\eps-\mc{I})$. We first explain how the $2\to 4$ branchings
are generated and how the integration variables are connected to the kinematic
variables introduced in App.~\ref{sec:phasespace}. The generation of endpoint
contributions is a simple extension of the generation of $2\to 4$ branchings,
and is described later on.

\subsubsection{Differential contributions}
The splitting kernels that are differential in the $2\to 4$ -particle phase space
are defined by the subtracted triple-collinear splitting functions of~\cite{Catani:1999ss}.
As shown in Sec.~\ref{sec:local_sub} we only need their four-dimensional values.
There are two independent flavor-changing contributions, which are given by
(cf.\ Eq.~\eqref{eq:pqqp_tc})
\begin{equation}\label{eq:fs_pqqp_tc}
  \begin{split}
    &\mr{R}_{qq'}^{(F)}(\tilde{z}_a,\tilde{z}_i,\tilde{z}_j,s_{ai},s_{aj},s_{ij})=\frac{1}{2}C_FT_R\frac{s_{aij}}{s_{ai}}
  \left[-\frac{t_{ai,j}^2}{s_{ai}s_{aij}}+\frac{4\,\tilde{z}_j+(\tilde{z}_a-\tilde{z}_i)^2}{\tilde{z}_a+\tilde{z}_i}
    +\left(\tilde{z}_a+\tilde{z}_i-\frac{s_{ai}}{s_{aij}}\right)\right]\;,\\
  &\mr{R}_{q\bar{q}}^{(F)}(\tilde{z}_a,\tilde{z}_i,\tilde{z}_j,s_{ai},s_{aj},s_{ij})=
  P_{qq'}^{1\to3}(\tilde{z}_a,\tilde{z}_i,\tilde{z}_j,s_{ai},s_{aj},s_{ij})\,\\
    &\qquad-\frac{1}{N_C}C_FT_R\frac{s_{aij}}{s_{ai}}\bigg\{\frac{2s_{ij}}{s_{aij}}
    +\frac{1+\tilde{z}_a^2}{1-\tilde{z}_i}-\frac{2\tilde{z}_i}{1-\tilde{z}_j}
    -\frac{s_{aij}}{s_{aj}}\frac{\tilde{z}_a}{2}\frac{1+\tilde{z}_a^2}{(1-\tilde{z}_i)(1-\tilde{z}_j)}
    \bigg\}+(i\leftrightarrow j)\;.
  \end{split}
\end{equation}
Their corresponding local subtraction terms are given by
\begin{equation}\label{eq:fs_pqqp_tc_2}
  \begin{split}
    \mr{S}_{qq'}^{(F)}(\tilde{z}_a,\tilde{z}_i,\tilde{z}_j,s_{ai},s_{aj},s_{ij})=&\;
    C_F T_R\frac{s_{aij}}{s_{ai}}\bigg[\,\frac{1+\tilde{z}_j^2}{1-\tilde{z}_j}
    \left(1-\frac{2\tilde{z}_a\tilde{z}_i}{(\tilde{z}_a+\tilde{z}_i)^2}\right)
    +\frac{4 \tilde{z}_a \tilde{z}_i \tilde{z}_j}{(1-\tilde{z}_j)^3}\left(1-2\cos^2\phi_{a,j}^{i,k}\right)\bigg]\;,\\
    \mr{S}_{q\bar{q}}^{(F)}(\tilde{z}_a,\tilde{z}_i,\tilde{z}_j,s_{ai},s_{aj},s_{ij})=&\;
    \mr{S}_{qq'}^{(F)}(\tilde{z}_a,\tilde{z}_i,\tilde{z}_j,s_{ai},s_{aj},s_{ij})+(i\leftrightarrow j)\;.
  \end{split}
\end{equation}
Note that the subtraction term for $P_{q\bar{q}}$ is the simple sum of two
subtraction terms for $P_{qq'}$, i.e.\ the interference contribution
on the last line of Eq.~\eqref{eq:fs_pqqp_tc} does not create a new singularity.
The fully differential initial-state $2\to 4$ splitting kernels are defined
by crossing (cf.\ Eq.~\eqref{eq:tcps_def_rs})
\begin{equation}\label{eq:is_pqqp_tc}
  \begin{split}
    \mr{R}^{(I)}(\tilde{z}_a,\tilde{z}_i,\tilde{z}_j,s_{ai},s_{aj},s_{ij})=&\;\tilde{z}_a\mr{R}^{(F)}
    (1/\tilde{z}_a,-\tilde{z}_i/\tilde{z}_a,-\tilde{z}_j/\tilde{z}_a,-s_{ai},-s_{aj},s_{ij})\;,\\
    \mr{S}^{(I)}(\tilde{z}_a,\tilde{z}_i,\tilde{z}_j,s_{ai},s_{aj},s_{ij})=&\;\tilde{z}_a\mr{S}^{(F)}
    (1/\tilde{z}_a,-\tilde{z}_i/\tilde{z}_a,-\tilde{z}_j/\tilde{z}_a,-s_{ai},-s_{aj},s_{ij})\;.
  \end{split}
\end{equation}

The kinematics for $2\to 4$ branchings in our parton-shower implementation
is described in App.~\ref{sec:phasespace}, and the kinematics for $2\to 3$
branchings can be found in~\cite{Hoche:2015sya}. For a numerical
implementation of Eqs.~\eqref{eq:fs_pqqp_tc}-\eqref{eq:is_pqqp_tc}
it is important to match the definition of splitting variables
in~\cite{Catani:1999ss}, or else the local cancellation of singularities
will fail. We describe in the following how these variables
are chosen in practice, based on the phase-space variables in
App.~\ref{sec:phasespace}. We note that in our Monte-Carlo implementations
all four-momenta of the $2\to4$ parton final state are known at the time
the splitting kernel is evaluated. We could therefore simply use the formal
definitions in~\cite{Catani:1999ss}. However, we find it instructive to
write the arguments of the splitting kernels explicitly in terms of the
variables used in App.~\ref{sec:phasespace}.

In the case of final-state emitter with final-state spectator, we have
the evolution and splitting variables (see App.~\ref{sec:css_kin_ff_123})
\begin{equation}
  t=\frac{4\,p_jp_{ai}\,p_{ai}p_k}{q^2}\;,\quad
  z_a=\frac{2\,p_ap_k}{q^2}\;
  \qquad\text{and}\qquad
  s_{ai}\;,\quad
  x_a=\frac{p_ap_k}{p_{ai}p_k}\;.
\end{equation}
We can thus identify the variables in Eqs.~\eqref{eq:fs_pqqp_tc}
and~\eqref{eq:fs_pqqp_tc_2} as follows
\begin{equation}
  \tilde{z}_a=\frac{z_a}{1-s_{aij}/q^2}\;,\qquad
  \tilde{z}_i=\frac{\xi_a-z_a}{1-s_{aij}/q^2}\;,\qquad
  \tilde{z}_j=1-\tilde{z}_a-\tilde{z}_i
  \qquad\text{where}\qquad
  s_{aij}=t/\xi_a+s_{ai}\;.\qquad
\end{equation}
The scalar products $s_{aj}$ and $s_{ij}$ are computed explicitly.
We transform the $s_{ai}$ integration such as to obtain a value
in the physical region $s_{ai}\le s_{aij}$.
\begin{equation}\label{eq:sai_v_ff}
  {\rm d}s_{ai}=\frac{{\rm d}\tilde{v}}{1-\tilde{v}}\,s_{aij}\;,
  \qquad\text{where}\qquad
  \tilde{v}=\frac{s_{ai}}{s_{aij}}\;.
\end{equation}
The factor $s_{aij}$ on the right hand side cancels one of the denominators
in the term in parentheses of Eq.~\eqref{eq:tc_me_factorization}.
In the case of final-state emitter with initial-state spectator, we have
the evolution and splitting variables (see App.~\ref{sec:css_kin_fi_123})
\begin{equation}
  t=\frac{2\,p_jp_{ai}\,p_{ai}p_b}{p_{aij}p_b}\;,\quad
  z_a=\frac{p_ap_b}{p_{aij}p_b}\;
  \qquad\text{and}\qquad
  s_{ai}\;,\quad
  x_a=\frac{p_ap_b}{p_{ai}p_b}\;
\end{equation}
We identify the variables in Eqs.~\eqref{eq:fs_pqqp_tc}
and~\eqref{eq:fs_pqqp_tc_2} as follows
\begin{equation}
  \tilde{z}_a=z_a\;,\qquad
  \tilde{z}_i=\xi_a-z_a\;,\qquad
  \tilde{z}_j=1-\xi_a\;.\qquad
\end{equation}
The scalar products $s_{aj}$ and $s_{ij}$ are computed explicitly, and
the $s_{ai}$ integration is transformed as in Eq.~\eqref{eq:sai_v_ff}.
In the case of initial-state emitter with final-state spectator, we have
the evolution and splitting variables (see App.~\ref{sec:css_kin_if_123})
\begin{equation}
  \begin{split}
    t=\frac{2\,p_jp_{ai}\,p_{ai}p_k}{p_ap_{ijk}}\;,\quad
    z_a=\frac{-q^2}{2\,p_ap_{ijk}}
    \qquad\text{and}\qquad
    s_{ai}\;,\quad
    x_a=\frac{p_{ai}p_k}{p_ap_{ijk}}\;.
  \end{split}
\end{equation}
We identify the variables in Eqs.~\eqref{eq:is_pqqp_tc} as follows
\begin{equation}
  \frac{-\tilde{z}_j}{\tilde{z}_a}=1-\frac{C}{\xi_a}\;,\qquad
  \frac{-\tilde{z}_i}{\tilde{z}_a}=\frac{2p_ip_k}{q^2/C}\;,\qquad
  \frac{1}{\tilde{z}_a}=1-\tilde{z}_i-\tilde{z}_j\;,
  \qquad\text{where}\qquad
  \frac{1}{C}=1+\frac{t/x_a-s_{ai}}{q^2}\;.\qquad
\end{equation}
The scalar products $s_{aj}$ and $s_{ij}$ are computed explicitly.
Using the relation $s_{aij}=-t/x_a+s_{ai}+m_j^2$, we transform
the $s_{ai}$ integration such as to obtain a value in
the physical region $s_{ai}\le s_{aij}/\xi_a$.
\begin{equation}\label{eq:sai_v_if}
  {\rm d}s_{ai}=\frac{{\rm d}\tilde{v}}{1-\tilde{v}/\xi}\,s_{aij}\;,
  \qquad\text{where}\qquad
  \tilde{v}=\xi_a\frac{s_{ai}}{s_{aij}}\;.
\end{equation}
Note that in this case the $\tilde{v}$ integral is limited to $0<\tilde{v}\le\xi_a$.
The factor $s_{aij}$ on the right hand side cancels one of the denominators
in the term in parentheses of Eq.~\eqref{eq:tc_me_factorization}.
In the case of initial-state emitter with initial-state spectator, we have
the evolution and splitting variables (see App.~\ref{sec:css_kin_ii_123})
\begin{equation}
  \begin{split}
    t=\frac{2\,p_jp_{ai}\,p_{ai}p_b}{p_ap_b}\;,\quad
    z_a=\frac{q^2}{2\,p_ap_b}
    \qquad\text{and}\qquad
    s_{ai}\;,\quad
    x_a=\frac{p_{ai}p_b}{p_ap_b}\;.
  \end{split}
\end{equation}
We identify the variables in Eqs.~\eqref{eq:is_pqqp_tc} as follows
\begin{equation}
  \frac{1}{\tilde{z}_a}=\frac{C}{z_a}\;,\qquad
  \frac{-\tilde{z}_i}{\tilde{z}_a}=-\frac{1-x_a}{z_a/C}\;,\qquad
  \frac{-\tilde{z}_j}{\tilde{z}_a}=1-\frac{C}{\xi_a}\;,
  \qquad\text{where}\qquad
  \frac{1}{C}=1+\frac{t/x_a-s_{ai}}{q^2}\;.\qquad
\end{equation}
The scalar products $s_{aj}$ and $s_{ij}$ are computed explicitly, and
the $s_{ai}$ integration is transformed as in Eq.~\eqref{eq:sai_v_if}.

We use the Sudakov veto algorithm to select the evolution variable $t$,
based on an overestimate that is given by the soft enhanced term of the
leading-order $q\to g$ splitting function. The variable $z_a$ is selected
accordingly, and the variable $x_a$ is generated logarithmically between
$z_a$ and 1. The variable $v$ is generated uniformly between 0 and 1.

Negative values of the splitting kernels are handled using the weighting
technique presented in~\cite{Hoeche:2009xc,Lonnblad:2012hz}. If we assume
for the moment that the splitting function is given by $f$, and we use
the overestimate $g$, then we can introduce an auxiliary overestimate 
$h$ which is adjusted such that the probability $f/h$ to accept a splitting
conforms to $f/h \in [0,1]$. This implies that $h$ may have a similarly
complex functional dependence on the phase-space variables as $f$ itself.
The fact that $f/h$ is used as accept probability in the Monte Carlo
implementation is corrected by a multiplicative weight, which
ensures the proper exponentiation of the desired branching probability.
\begin{equation}
w = \frac{h}{g}\times
\begin{cases}
~\displaystyle \frac{g-f}{h-f} & \quad{\textnormal{if the splitting was rejected,}}\\[3mm]
~\displaystyle 1               & \quad{\textnormal{if the splitting was accepted.}}
\end{cases}
\end{equation}

\subsubsection{Endpoint contributions}
In order to implement Eq.\eqref{eq:tcps_mc} in a parton shower, we find it convenient
to perform the integration of $(\mr{I}+\mc{P}/\eps-\mc{I})$ numerically using the method
outlined in Sec.~\ref{sec:ps_formalism}. This will eventually allow us to match the
phase-space coverage of the real-correction and the local subtraction terms in the
corresponding integrated MC counterterms. Note that the phase-space coverage is restricted
in the region $t>0$, as the $z_a$ integration range is limited by momentum conservation,
cf.\ Sec.~\ref{sec:ps_formalism}.
This phase-space restriction is the main difference between the algorithm proposed here and
the analytic calculation in Sec.~\ref{sec:pqq_analytic}.
In addition, a fully numerical evaluation of $(\mr{I}+\mc{P}/\eps-\mc{I})$ allows us
to extend the calculation to splitting functions that we have not previously computed
analytically, such as the flavor-changing contributions $P_{q\bar{q}}$. Note that this
kernel in particular does not require any new endpoint contributions beyond those that
can be obtained from crossing relations. Thus, the full benefit of our method will become
apparent only when implementing the more complicated triple-collinear splitting functions.

The procedure for the MC integration of $(\mr{I}+\mc{P}/\eps-\mc{I})$ is as follows: We generate
configurations in the $2\to4$-parton phase space as described in App.~\ref{sec:phasespace},
which are subsequently projected onto $s_{ai}=0$, while the dependence on $x_a$ and $\xi_a$
is retained. This corresponds to singling out the pole term in the expansion
\begin{equation}
  \frac{1}{v^{1+\eps}}=-\frac{1}{\eps}\,\delta(v)+\sum_{i=0}^\infty\frac{\eps^n}{n!}\left(\frac{\log^n v}{v}\right)_+
\end{equation}
The $1/\eps$ poles that are generated in this manner will cancel between the integrated
subtraction term, $\mr{I}$, and the renormalization term, $\mc{P}$. In order to compute
the finite remainder of $(\mr{I}+\mc{P}/\eps-\mc{I})$, we simply need to implement the $\mc{O}(\eps)$
terms in the expansion of the {\it differential} forms of the subtraction and matching terms.
They are given by
\begin{equation}\label{eq:iterm_mc}
  \begin{split}
    \Delta\mr{I}_{qq'}^{(F)}(\tilde{z}_a,\tilde{z}_i,\tilde{z}_j)=&\;
    \tilde{\mr{I}}_{qq'}(\tilde{z}_a,\tilde{z}_i,\tilde{z}_j,\tilde{z}_a)-
    \tilde{\mc{I}}_{qq'}(\tilde{z}_a,\tilde{z}_i,\tilde{z}_j,\tilde{z}_a+\tilde{z}_i)\;,\\
    \Delta\mr{I}_{qq'}^{(I)}(\tilde{z}_a,\tilde{z}_i,\tilde{z}_j)=&\;
    \tilde{z}_a\left[\,\tilde{\mr{I}}_{qq'}\Big(\frac{1}{\tilde{z}_a},
      \frac{-\tilde{z}_i}{\tilde{z}_a},\frac{-\tilde{z}_j}{\tilde{z}_a},
      \frac{\tilde{z}_a}{\tilde{z}_a+\tilde{z}_j}\Big)
    -\tilde{\mc{I}}_{qq'}\Big(\frac{1}{\tilde{z}_a},
    \frac{-\tilde{z}_i}{\tilde{z}_a},\frac{-\tilde{z}_j}{\tilde{z}_a},
    \frac{-\tilde{z}_a}{\tilde{z}_a+\tilde{z}_j}\Big)\,\right]\;.
  \end{split}
\end{equation}
where
\begin{equation}
  \begin{split}
    \tilde{\mr{I}}_{qq'}(\tilde{z}_a,\tilde{z}_i,\tilde{z}_j,\tilde{x})=&\;
    C_F T_R \left[\frac{1+\tilde{z}_j^2}{1-\tilde{z}_j}
      +\left(1-\frac{2\,\tilde{z}_a\tilde{z}_i}{(\tilde{z}_a+\tilde{z}_i)^2}\right)
      \left(1-\tilde{z}_j+\frac{1+\tilde{z}_j^2}{1-\tilde{z}_j}\right)
      \Big(\log(\tilde{x}\,\tilde{z}_i\tilde{z}_j)-1\Big)\right]\;,\\
    \tilde{\mc{I}}_{qq'}(\tilde{z}_a,\tilde{z}_i,\tilde{z}_j,\tilde{x})=&\;
    2C_F\left[\,\frac{1+\tilde{z}_j^2}{1-\tilde{z}_j}
      \log(\tilde{x}\,\tilde{z}_j)+(1-\tilde{z}_j)\,\right]
    P_{gq}^{(0)}\Big(\frac{\tilde{z}_a}{\tilde{z}_a+\tilde{z}_i}\Big)\;.
  \end{split}
\end{equation}
The endpoint contributions for $q\to\bar{q}$ transitions are obtained as a
sum of two terms of $q\to q'$ type
\begin{equation}
  \Delta\mr{I}_{q\bar{q}}(\tilde{z}_a,\tilde{z}_i,\tilde{z}_j)=
  \Delta\mr{I}_{qq'}(\tilde{z}_a,\tilde{z}_i,\tilde{z}_j)+(i\leftrightarrow j)\;.
\end{equation}

\subsubsection{Symmetry factors}
Finally, according to Sec.~\ref{sec:ps_formalism}, we multiply each term
in Eq.~\eqref{eq:tcps_mc} by an additional factor $z_a$ in branchings
with final-state emitter, independent of the type of spectator. This can be
interpreted as an identification of the parton for which the evolution
equation is constructed. The extension to $1\to 3$ splittings requires
a similar factor for one of the two radiated partons, if the two are
indistinguishable. In the case of the simulation presented here this
applies to the flavor-changing splittings of type $q\to\bar{q}$.
The corresponding extension of the symmetry relation,
Eq.~\eqref{eq:lokernels_symmetry}, reads
\begin{equation}\label{eq:nlokernels_symmetry}
  \begin{split}
    &\sum_{b=q,g}\int_{0}^{1-\eps} {\rm d}z_1
    \int_{0}^{1-\eps} {\rm d}z_2\,\frac{z_1\,z_2}{1-z_1}\,
    \Theta(1-z_1-z_2)\,P_{a\to ab\bar{b}}(z_1,z_2,\ldots)\\
    &\qquad=\sum_{b=q,g}\int_{\eps}^{1-\eps} {\rm d}z_1\,
    \int_{\eps}^{1-z_1} {\rm d}z_2\,S_{ab\bar{b}}\,P_{a\to ab\bar{b}}(z_1,z_2,\ldots)
    +\mc{O}(\eps)\;,\\
    &\sum_{\substack{b=q,g\\b\neq a}}\int_{0}^{1-\eps} {\rm d}z_1
    \int_{0}^{1-\eps} {\rm d}z_2\,\frac{z_1\,z_2}{1-z_1}\,\Theta(1-z_1-z_2)\,
    \Big(P_{a\to ba\bar{b}}(z_1,z_2,\ldots)+P_{a\to b\bar{b}a}(z_1,z_2,\ldots)\Big)\\
    &\qquad=\sum_{\substack{b=q,g\\b\neq a}}\int_{\eps}^{1-\eps} {\rm d}z_1\,
    \int_{\eps}^{1-z_1} {\rm d}z_2\,S_{ab\bar{b}}\,P_{a\to ba\bar{b}}(z_1,z_2,\ldots)
    +\mc{O}(\eps)\;,
  \end{split}
\end{equation}
where $S_{ab\bar{b}}=1/(\prod_{c=q,g}n_c!)$, with $n_c$ the number of partons
of type $c$, is the usual symmetry factor for the final-state $ab\bar{b}$.
Thus, all terms in Eq.~\eqref{eq:tcps_mc} are multiplied by the
following overall symmetry factors:
\begin{equation}
  \begin{split}
    S^{(F)}=&\;z_a\,\frac{\xi_a-z_a}{1-z_a}\;,\qquad
    &S^{(I)}=&\;\frac{1-x_a}{1-z_a}\;.
  \end{split}
\end{equation}

\section{Numerical results}
\label{sec:results}
\begin{figure}[p]
  \subfigure{
    \begin{minipage}{0.475\textwidth}
      \begin{center}
        \includegraphics[scale=0.7]{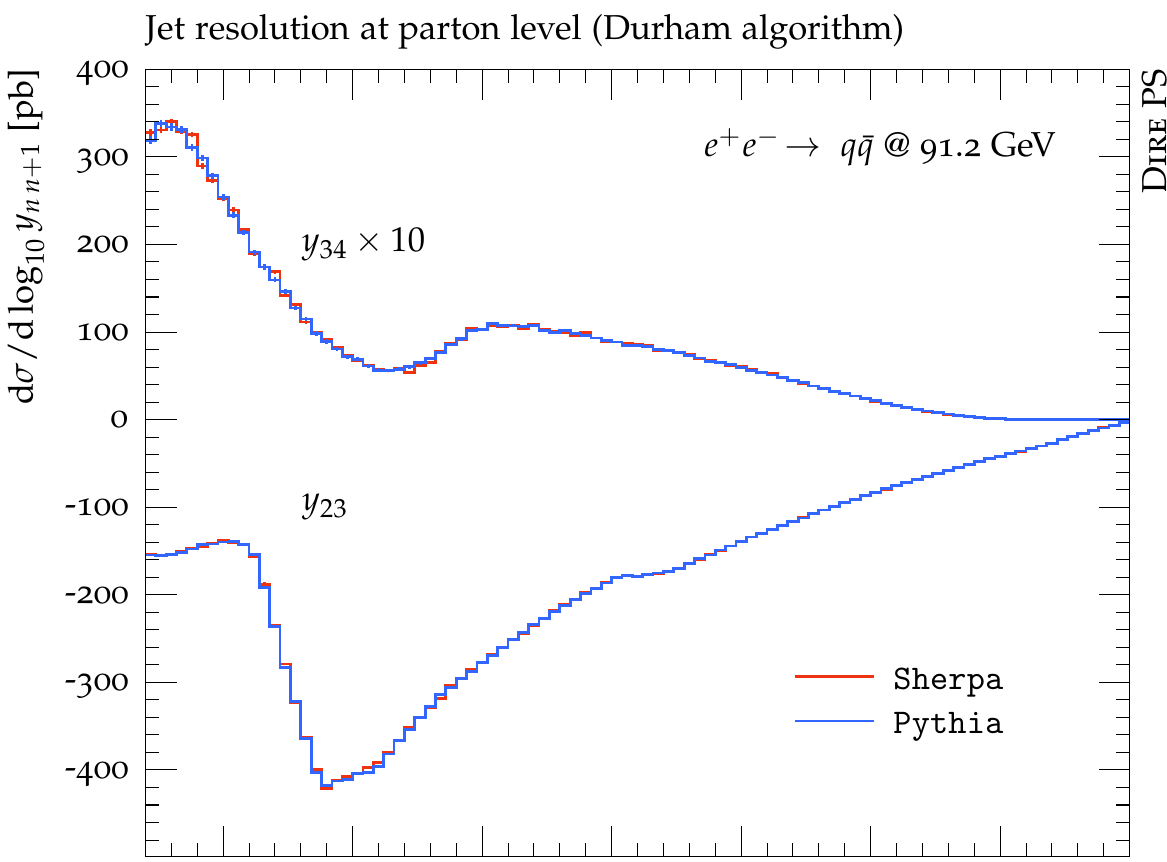}\\[-0.75mm]
        \includegraphics[scale=0.7]{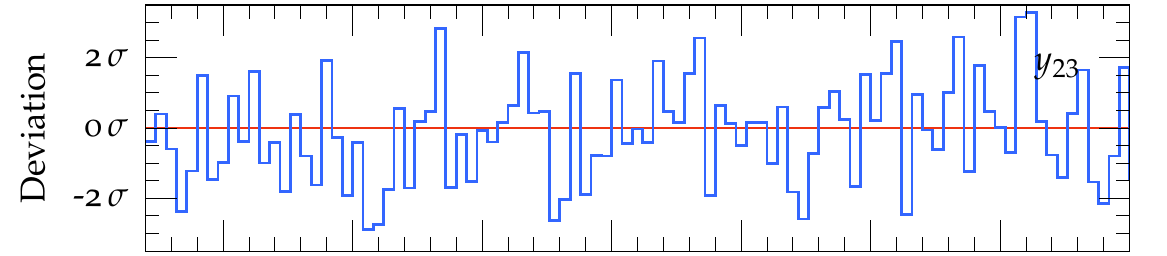}\\[-0.75mm]
        \includegraphics[scale=0.7]{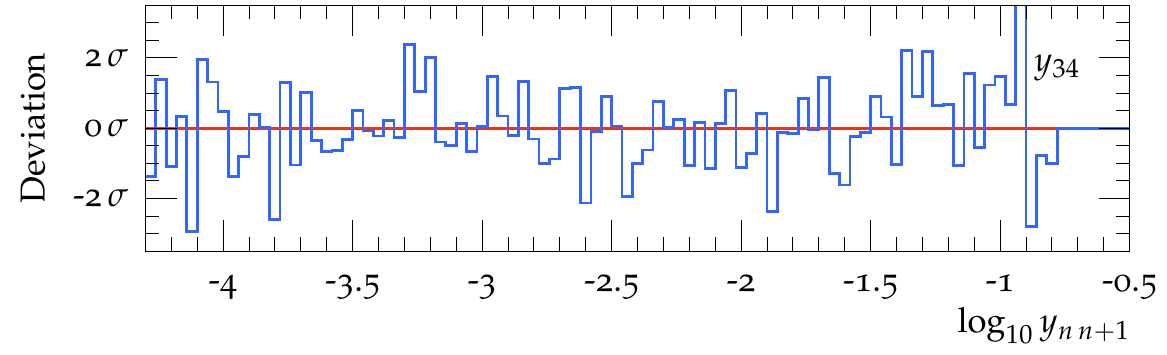}
      \end{center}
    \end{minipage}
    \label{fig:validation_ff}}
  \subfigure{
    \begin{minipage}{0.475\textwidth}
      \begin{center}
        \includegraphics[scale=0.7]{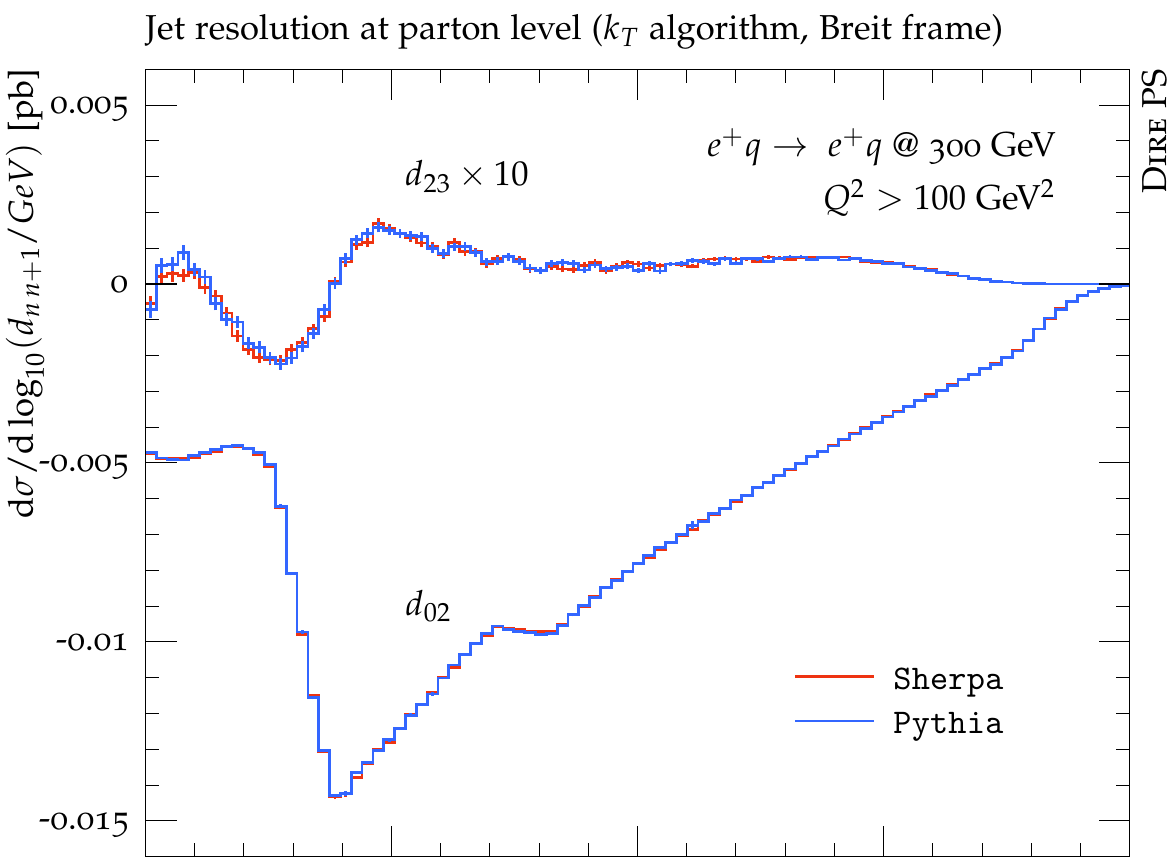}\\[-0.75mm]
        \includegraphics[scale=0.7]{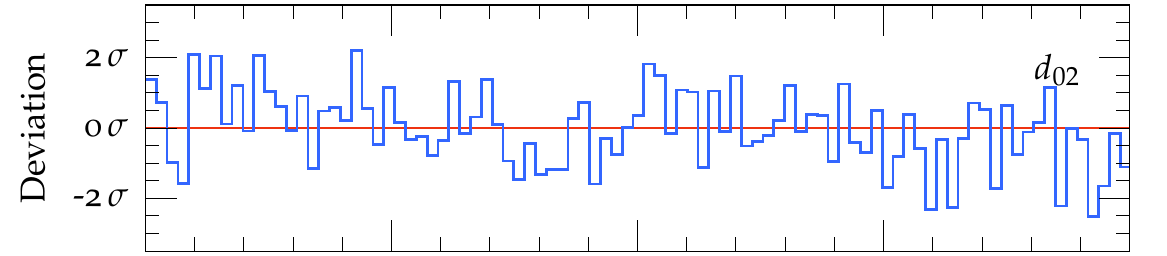}\\[-0.75mm]
        \includegraphics[scale=0.7]{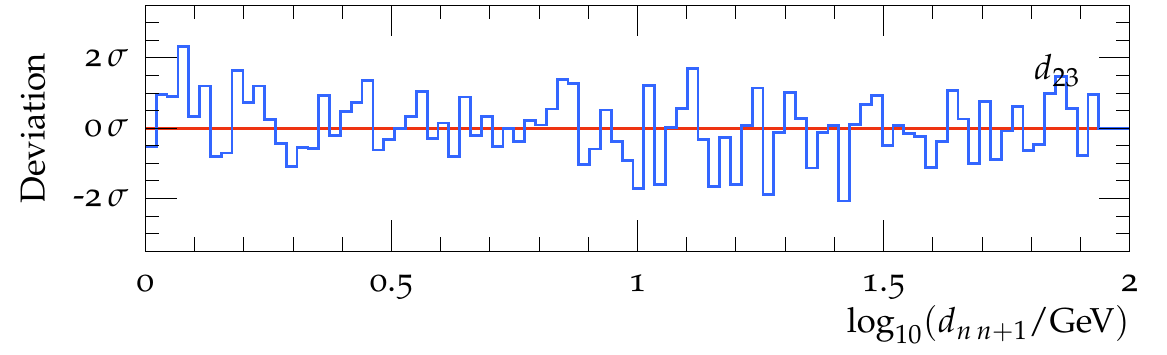}
      \end{center}
    \end{minipage}
    \label{fig:validation_fi}}
  \subfigure{
    \begin{minipage}{0.475\textwidth}
      \begin{center}
        \includegraphics[scale=0.7]{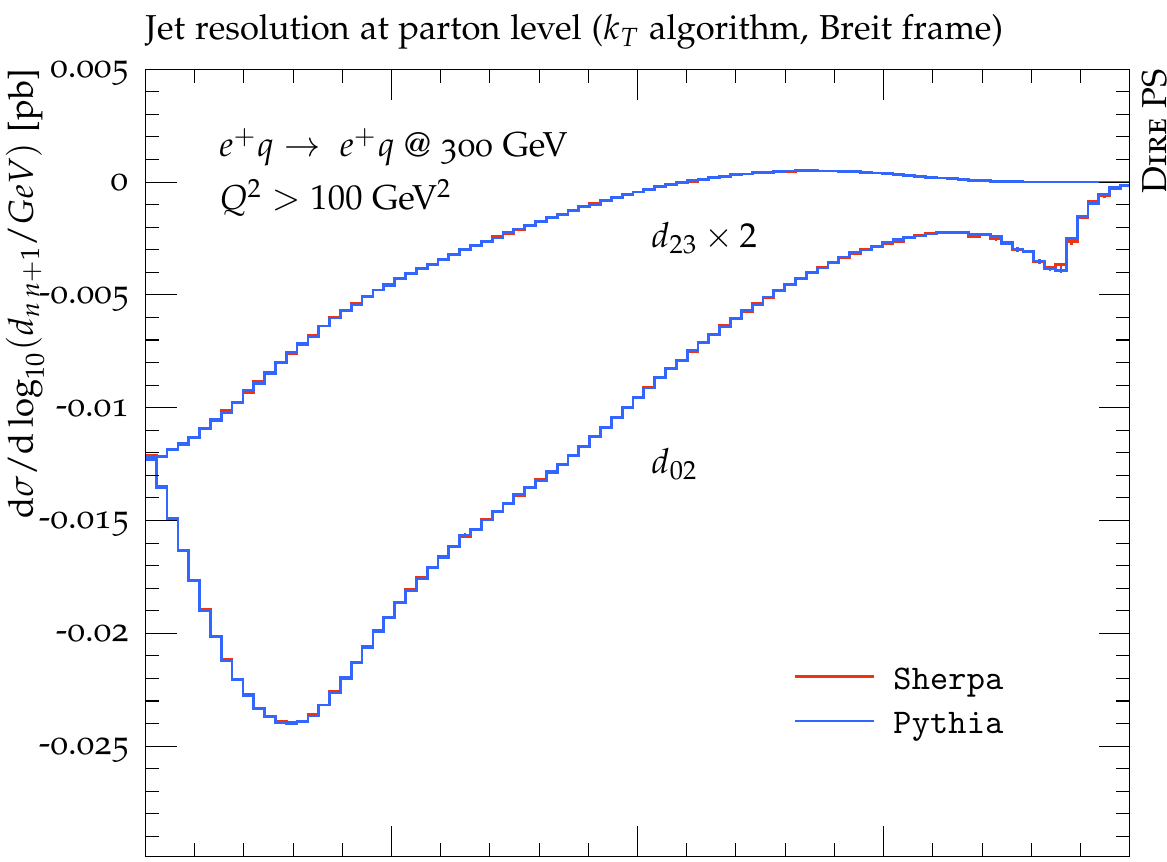}\\[-0.75mm]
        \includegraphics[scale=0.7]{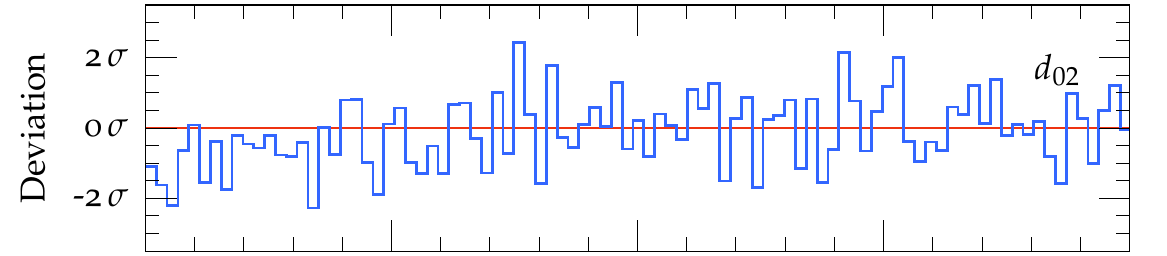}\\[-0.75mm]
        \includegraphics[scale=0.7]{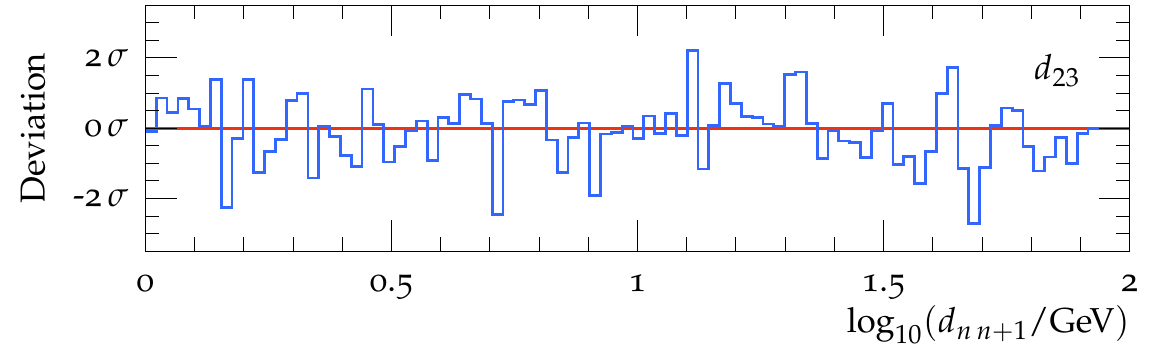}
      \end{center}
    \end{minipage}
    \label{fig:validation_if}}
  \subfigure{
    \begin{minipage}{0.475\textwidth}
      \begin{center}
        \includegraphics[scale=0.7]{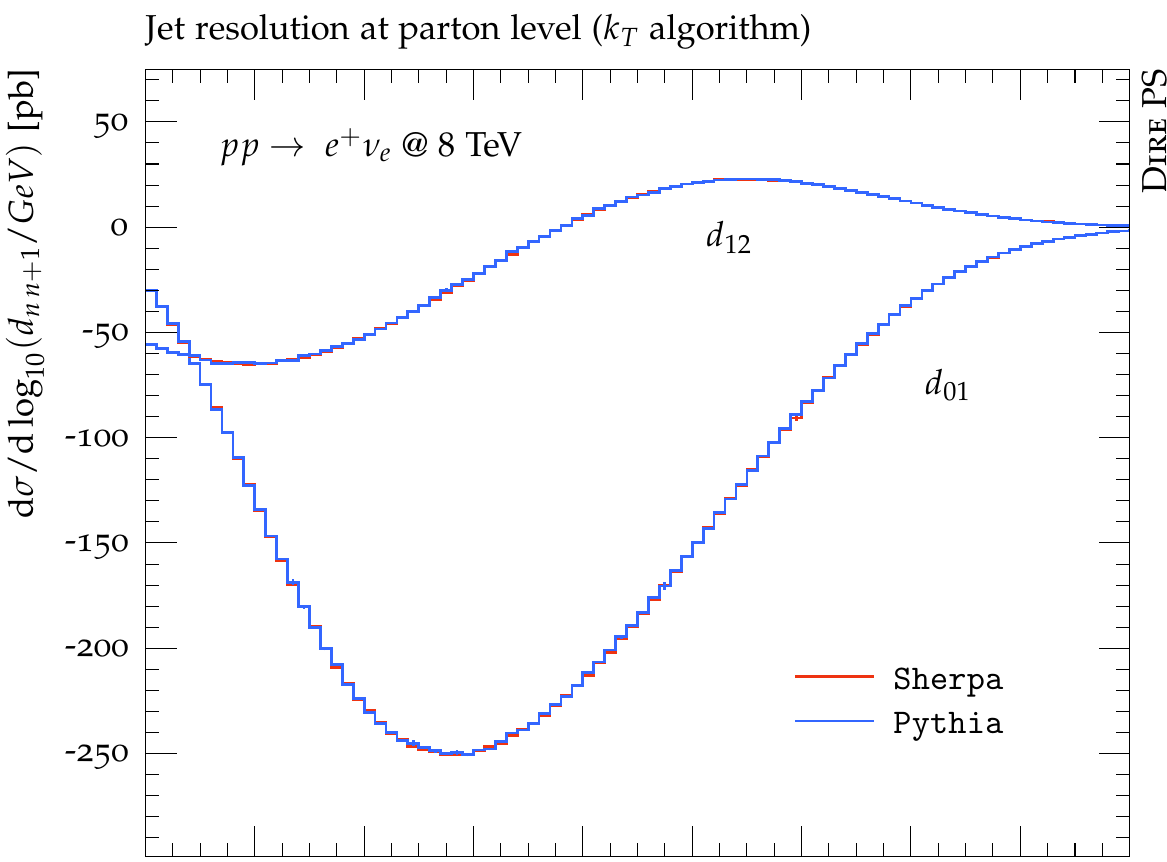}\\[-0.75mm]
        \includegraphics[scale=0.7]{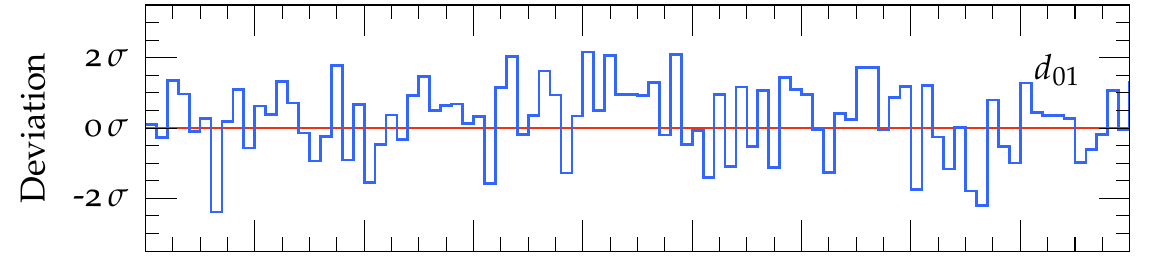}\\[-0.75mm]
        \includegraphics[scale=0.7]{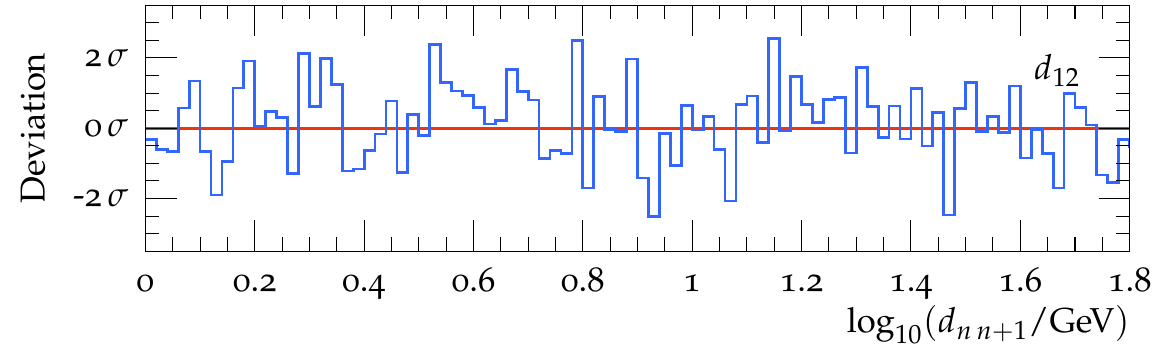}
      \end{center}
    \end{minipage}
    \label{fig:validation_ii}}
  \caption{Validation of the simulation of triple-collinear parton splittings
    in final-state (top row) and initial-state (bottom row) branchings with
    final-state (left panels) and initial-state (right panels) spectator.
    We show Durham $k_T$-jet rates in $e^+e^-\to$hadrons at LEP I,
    $k_T$-jet rates in neutral current DIS at HERA II with $Q^2>100~{\rm GeV}^2$,
    and $k_T$-jet rates in $pp\to e^+\nu_e$ at the 8~TeV LHC (top left to bottom right).
    \label{fig:validation}}
\end{figure}
\begin{figure}[p]
  \subfigure{
    \begin{minipage}{0.475\textwidth}
      \begin{center}
        \includegraphics[scale=0.7]{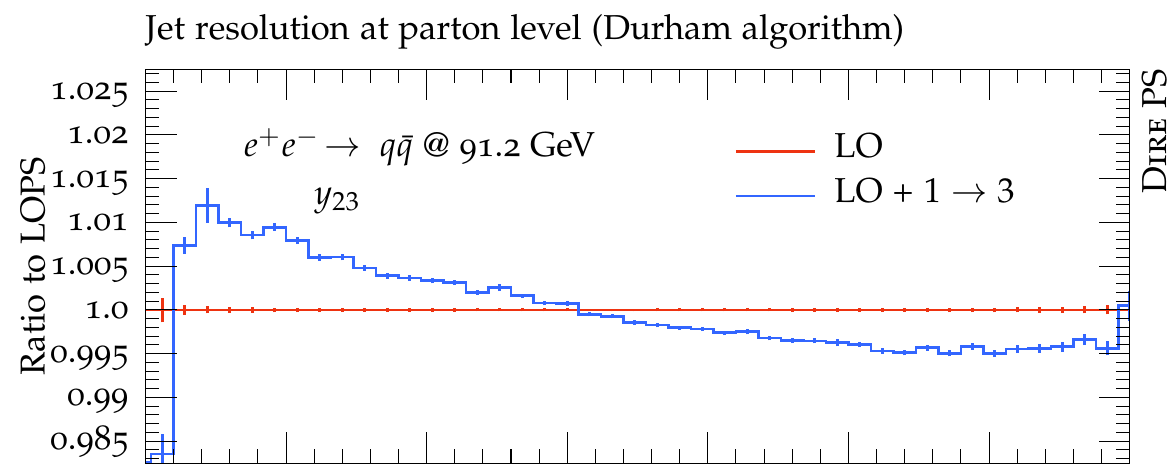}\\[-0.5mm]
        \includegraphics[scale=0.7]{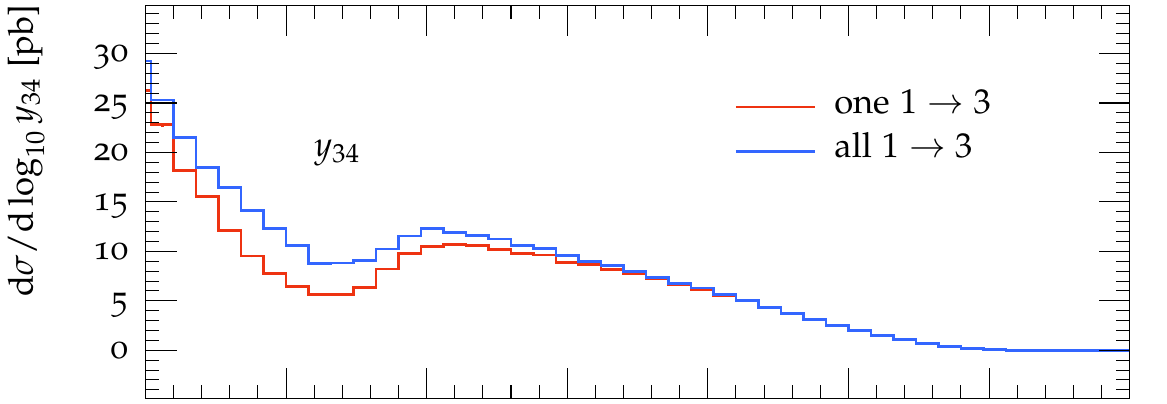}\\[-0.5mm]
        \includegraphics[scale=0.7]{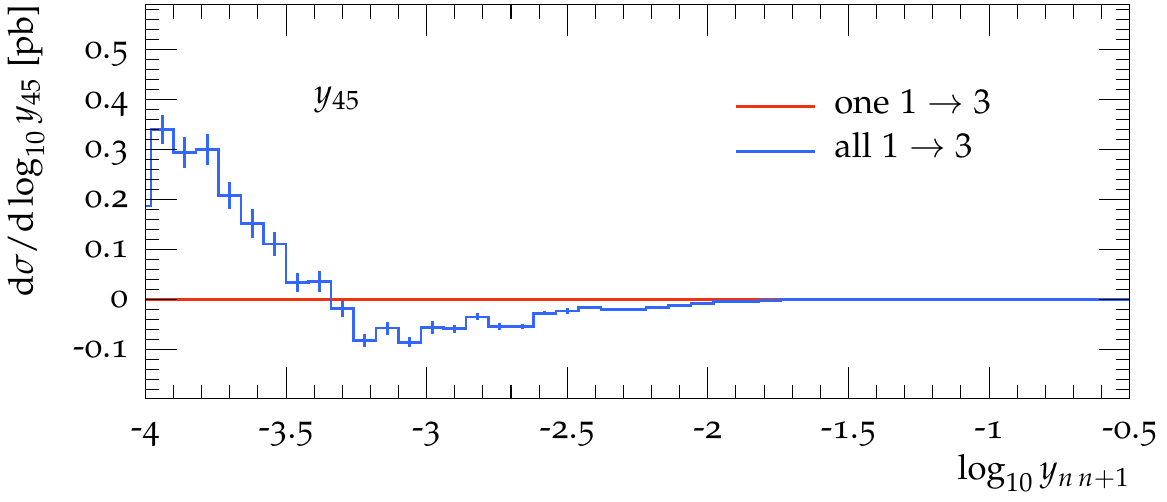}
      \end{center}
    \end{minipage}
    \label{fig:impact_fi}}
  \subfigure{
    \begin{minipage}{0.475\textwidth}
      \begin{center}
        \includegraphics[scale=0.7]{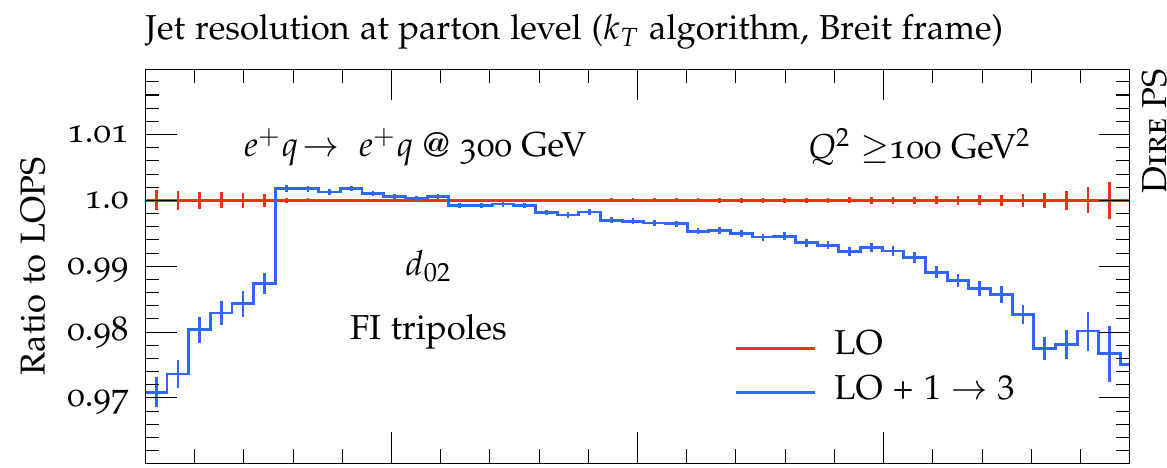}\\[-0.5mm]
        \includegraphics[scale=0.7]{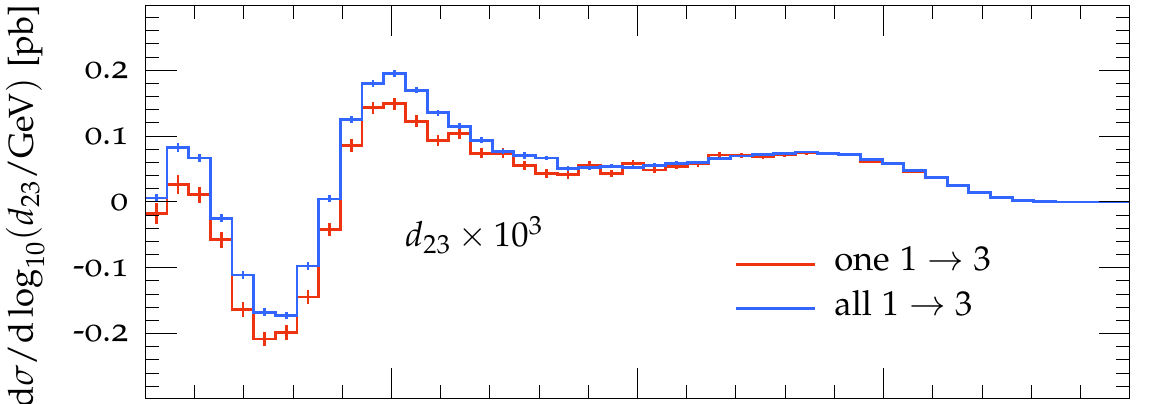}\\[-0.5mm]
        \includegraphics[scale=0.7]{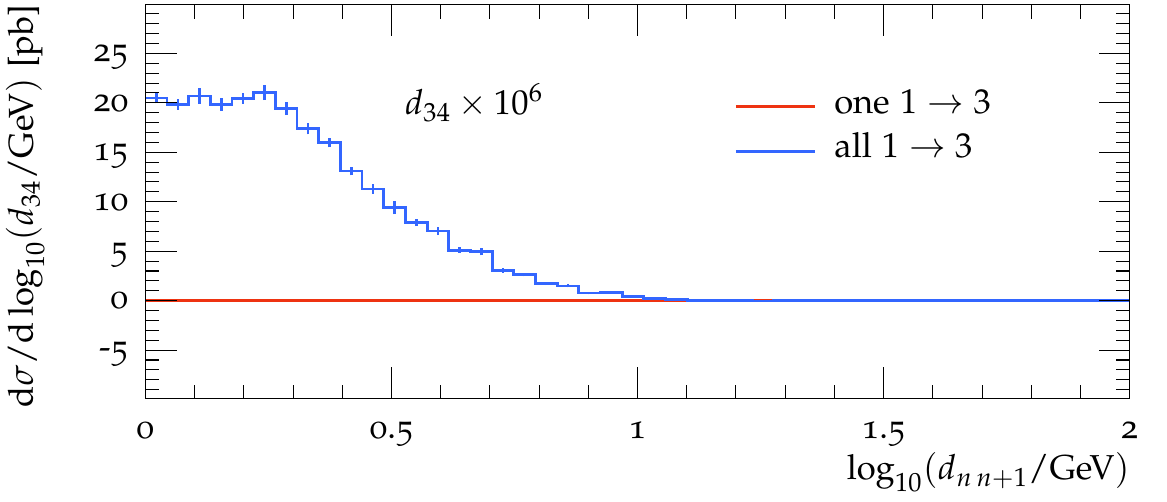}
      \end{center}
    \end{minipage}
    \label{fig:impact_ff}}
  \subfigure{
    \begin{minipage}{0.475\textwidth}
      \begin{center}
        \includegraphics[scale=0.7]{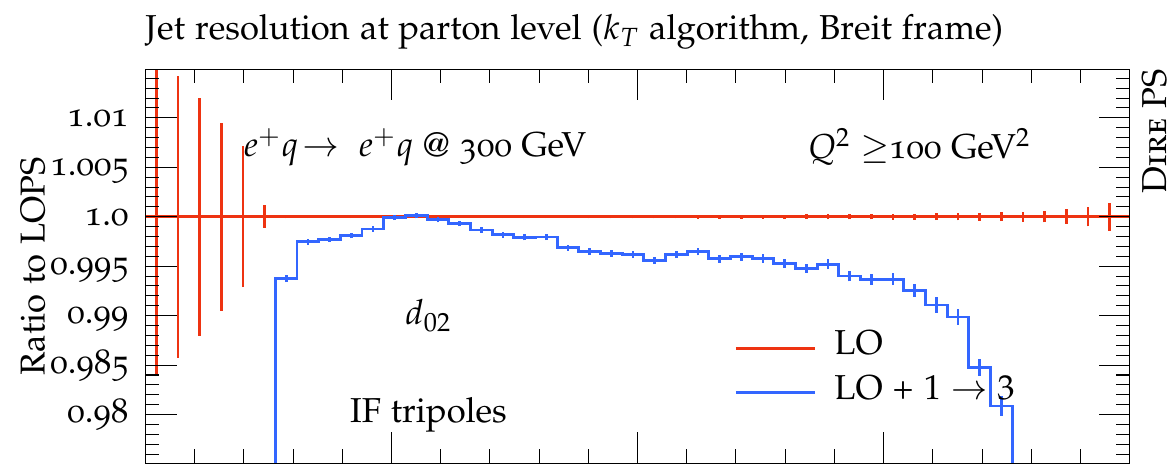}\\[-0.5mm]
        \includegraphics[scale=0.7]{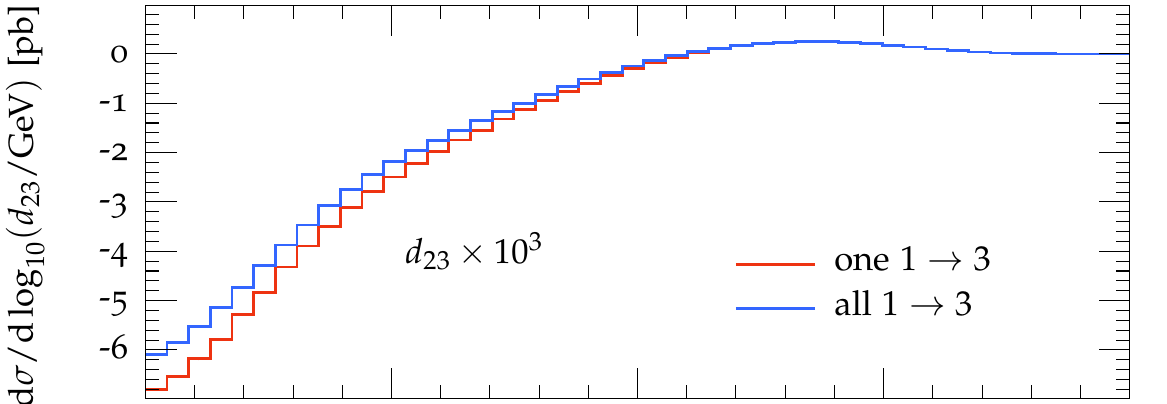}\\[-0.5mm]
        \includegraphics[scale=0.7]{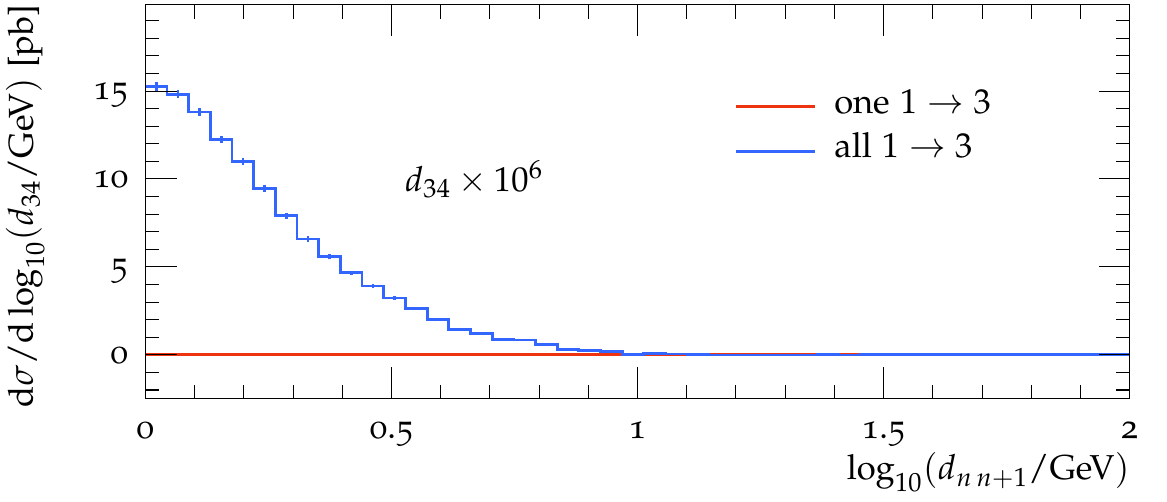}
      \end{center}
    \end{minipage}
    \label{fig:impact_if}}
  \subfigure{
    \begin{minipage}{0.475\textwidth}
      \begin{center}
        \includegraphics[scale=0.7]{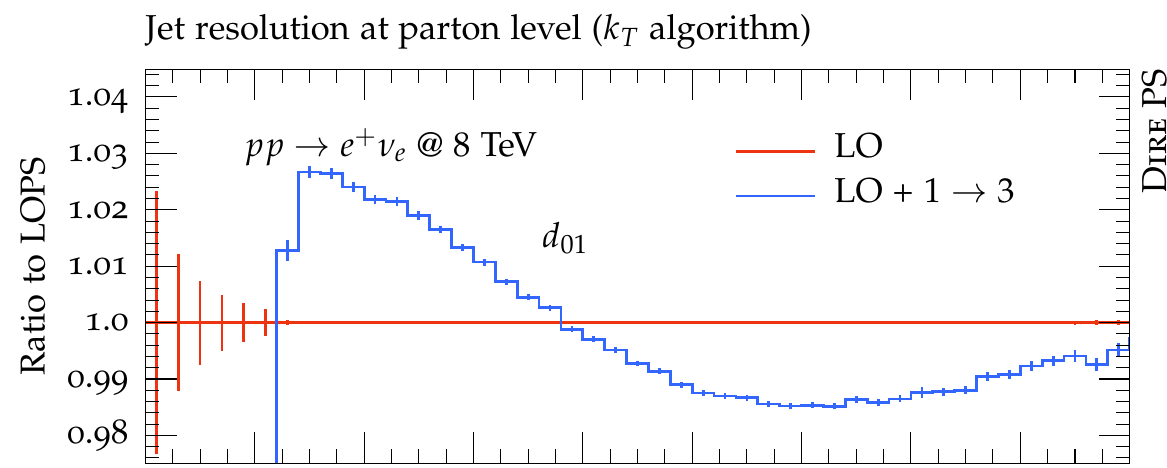}\\[-0.5mm]
        \includegraphics[scale=0.7]{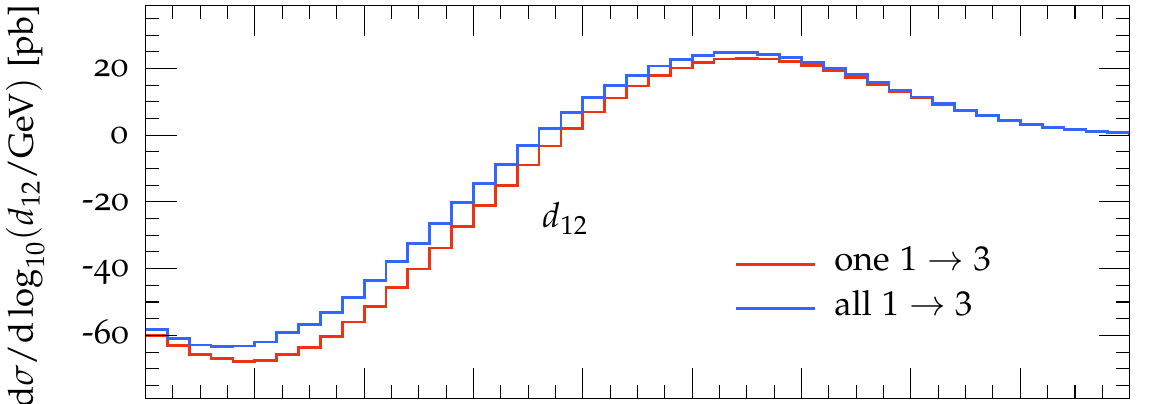}\\[-0.5mm]
        \includegraphics[scale=0.7]{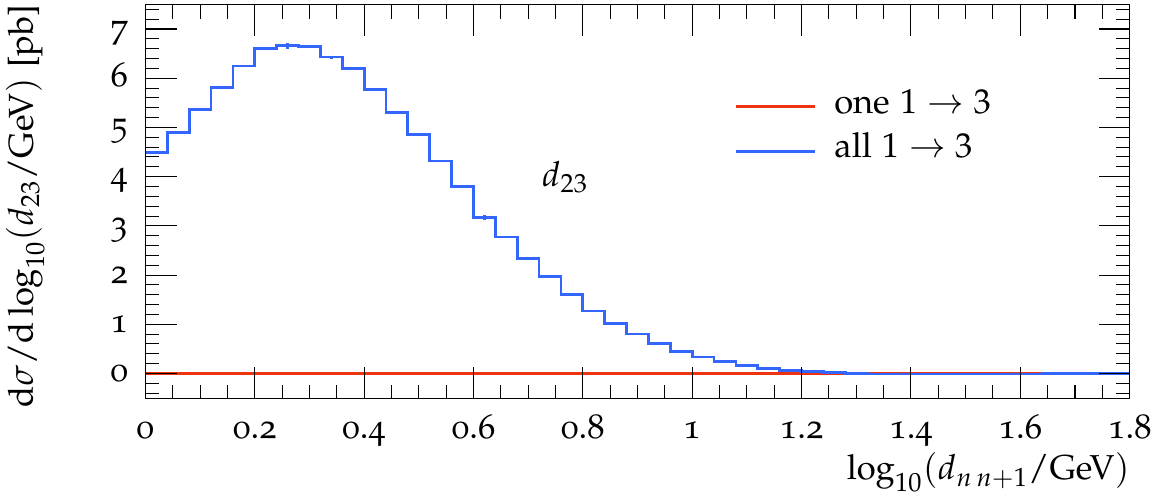}
      \end{center}
    \end{minipage}
    \label{fig:impact_ii}}
  \caption{Impact of the simulation of triple-collinear parton splittings
    on final-state (top row) and initial-state (bottom row) evolution with
    final-state (left panels) and initial-state (right panels) spectator.
    Top panels show the ratio between the leading-order result and the
    leading-order simulation including triple-collinear branchings. Middle
    and bottom panels show a comparison between the simulation of up to one
    triple-collinear splitting and arbitrarily many (both not including
    the leading-order result). For details, see Fig.~\ref{fig:validation}.
    \label{fig:impact}}
\end{figure}
In this section we present numerical cross-checks of our algorithm, and we compare the
magnitude of the corrections generated by the flavor-changing triple-collinear splitting
functions to the leading-order parton-shower result. We have implemented our algorithm
into the \Dire parton showers, which implies two entirely independent realizations within
the general purpose event generation frameworks \Pythia~\cite{Sjostrand:1985xi,Sjostrand:2014zea}
and \Sherpa~\cite{Gleisberg:2003xi,Gleisberg:2008ta}.
We employ the CT10nlo PDF set~\cite{Lai:2010vv}, and use the corresponding form
of the strong coupling. Following standard practice to improve the logarithmic accuracy
of the parton shower, the soft enhanced term of the leading-order splitting functions is rescaled
by $1+\alpha_s(t)/(2\pi) K$, where $K=(67/18-\pi^2/6)\,C_A-10/9\,T_R\,n_f$~\cite{Catani:1990rr}.

Figure~\ref{fig:validation} shows comparisons between the results from \Dire{}+\Pythia
and \Dire{}+\Sherpa for a single triple collinear splitting. Each simulation contains $10^9$ events.
The lower panels present the deviation between the two predictions, normalized to the statistical
uncertainty of \Dire{}+\Sherpa in the respective bin. If both implementations are equivalent, this
distribution should exhibit statistical fluctuations only. We validate final-state emissions with
final-state spectator in the reaction $e^+e^-\to\text{hadrons}$ (Fig.~\ref{fig:validation_ff}),
final-state emissions with initial-state spectator and initial-state emissions with final-state spectator
in the reaction $e^+p\to e^+\text{jet}$ (Figs.~\ref{fig:validation_fi} and~\ref{fig:validation_if}),
and initial-state emissions with initial-state spectator in the reaction $pp\to e^+\nu_e$ 
(Fig.~\ref{fig:validation_ii}). As required, the two implementations agree perfectly.
Each panel shows the predictions for the leading two differential jet rates, which are both
populated by the simulation of a single triple collinear parton branching. Note that their
numerical values can be both positive and negative, since the triple collinear splitting functions
are not positive-definite.
While the sub-leading jet rate receives contributions from the simulation of $\mr{R}-\mr{S}$
in Eq.~\eqref{eq:tcps_mc} only, the leading jet rate also receives contributions from $\mr{I}-\mc{I}$.
It can be seen that in all cases $\mr{I}-\mc{I}$ is much large on average than $\mr{R}-\mr{S}$.
The feature around -2.5 in Fig.~\ref{fig:validation_ff} and around 0.7 in Fig.~\ref{fig:validation_fi}
is due to the onset of $b$-quark production, which we include in the simulation only if $t>m_b^2$.
Similar, yet less pronounced, features are present in Figs.~\ref{fig:validation_if} and~\ref{fig:validation_ii}.

Figure~\ref{fig:impact} shows the impact of triple-collinear parton branchings
on the full evolution. We compare the ratio of leading-jet rates with and without
the simulation of $1\to 3$ splittings (upper panels), and we analyze the impact
of multiple $1\to 3$ splittings compared to a single one (middle and lower panels).
The edge in the ratio plots is related to the parton-shower cutoff, where the
$1\to 3$ splittings have a different behavior compared to the leading-order ones
due to the different evolution variable. It is apparent that the effect of
multiple triple-collinear branchings is marginal, even more so when compared to
the leading-order results, which are by themselves much larger in magnitude
than the correction from a single $1\to 3$ branching. We note again that the
largest part of the $1\to 3$ results is due to the subtraction terms $\mc{I}_{qq'}$,
which is in fact a leading-order like contribution.
The impact of the $1\to 3$ flavor-changing splittings is particularly small 
for $e^+e^-\to\text{hadrons}$. For $e^+p\to e^+\text{jet}$ scatterings, the 
hard-emission regions show the largest impact, while for $pp\to e^+\nu_e$, the
soft- and collinear-emission regions are enhanced.

\section{Conclusions}
\label{sec:conclusions}
We have presented a new scheme to include triple collinear splitting functions
into parton showers. As a proof of principle we have recomputed the timelike
and spacelike flavor-changing NLO DGLAP kernels $P_{qq'}$ and matched each
component of the integrand to the relevant parton-shower expression.
The implementation into two entirely independent Monte-Carlo simulations,
based on the general-purpose event generation frameworks \Pythia and \Sherpa
has been cross-checked to very high numerical accuracy. The impact of the
flavor changing triple-collinear kernels $P_{qq'}$ and $P_{q\bar{q}}$
has been studied in timelike and spacelike parton evolution as a first application.
We find that the numerical impact of the kernels investigated
here is marginal, with effects of up to $\sim 1$\% on differential jet rates
in $e^+e^-\to$hadrons at $\sqrt{s}$=91.2~GeV (LEP~I), neutral-current DIS with
$Q^2>100$~GeV$^2$ at $\sqrt{s}$=300~GeV (HERA~II), and $pp\to e^+\nu_e$ at 8~TeV (LHC~I).

\begin{acknowledgments}
  \noindent
  We thank Lance Dixon, Falko Dulat, Thomas Gehrmann, Frank Krauss,
  Silvan Kuttimalai and Leif L{\"o}nnblad for numerous fruitful discussions.
  This work was supported by the US Department of Energy
  under contracts DE--AC02--76SF00515 and DE--AC02--07CH11359.
\end{acknowledgments}

\appendix
\section{Kinematics and phase-space factorization for \boldmath{$1\to 3$} splittings}
\label{sec:phasespace}
In this section we give the phase-space parametrizations employed in our implementation
of $1\to 3$ parton branchings. We construct kinematic mappings that allow us to relate
the splitting and evolution variables to manifestly Lorentz invariant quantities.
In order to cover all possible applications, we list formulae for arbitrary external
particle masses. While this is not strictly needed in the course of this work, it may be
useful to include higher-order effects involving heavy quark splitting functions
in the future. The main results are Eqs.~\eqref{eq:tc_ff_ps}, \eqref{eq:tc_fi_ps},
\eqref{eq:tc_if_ps} and~\eqref{eq:tc_ii_ps}, as well as the corresponding $D$-dimensional
phase-space factors, Eqs.~\eqref{eq:dcdc_ps_fs} and~\eqref{eq:dcdc_ps_is}.

\subsection{Final-state emitter with final-state spectator}
\label{sec:css_kin_ff_123}
\begin{figure}
  \includegraphics[scale=1]{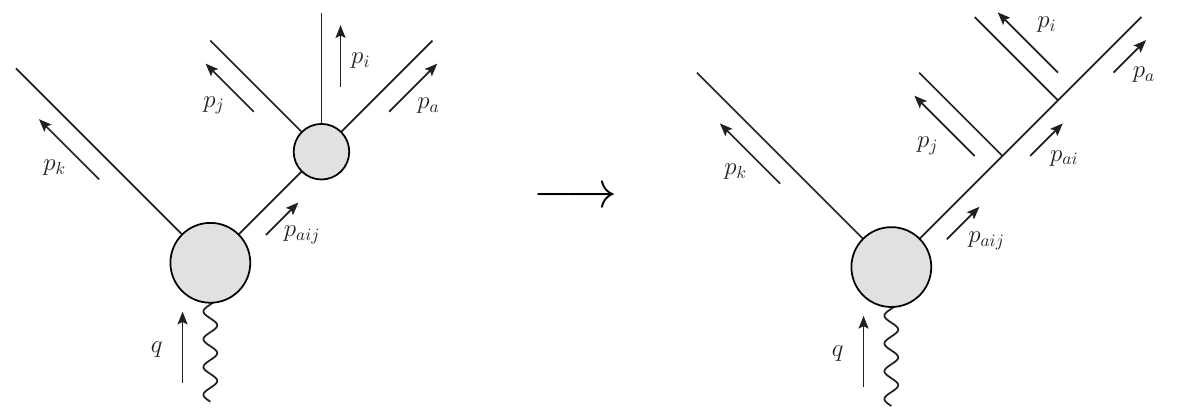}
  \caption{Kinematics mapping for final-state splittings with final-state spectator.
    \label{fig:kin_ff_123}}
\end{figure}

The kinematics for the case of a final-state radiator with final-state
spectator are derived from an iteration of the massive dipole kinematics
in~\cite{Catani:2002hc}. This is sketched in Fig.~\ref{fig:kin_ff_123}.
The evolution and splitting variables are defined as
\begin{equation}
  t=\frac{4\,p_jp_{ai}\,p_{ai}p_k}{q^2-m_{aij}^2-m_k^2}\;,\quad
  z_a=\frac{2\,p_ap_k}{q^2-m_{aij}^2-m_k^2}\;
  \qquad\text{and}\qquad
  s_{ai}\;,\quad
  x_a=\frac{p_ap_k}{p_{ai}p_k}\;.
\end{equation}
We generate the first branching $(\widetilde{a\imath\jmath},\tilde{k})\to(ai,j,k)$
with the mass of the pseudoparticle $ai$ set to the virtuality $s_{ai}$.
The new momentum of the spectator parton $k$ is determined as
\begin{equation}\label{eq:def_ff_pk}
  \begin{split}
    p_k^{\,\mu}=&\;\left(\tilde{p}_k^{\,\mu}-
    \frac{q\cdot\tilde{p}_k}{q^2}\,q^\mu\right)\,
    \sqrt{\frac{\lambda(q^2,s_{aij},m_k^2)}{\lambda(q^2,m_{aij}^2,m_k^2)}}
    +\frac{q^2+m_k^2-s_{aij}}{2\,q^2}\,q^\mu\;,
  \end{split}
\end{equation}
with $q=\tilde{p}_{aij}+\tilde{p}_k$ and $\lambda$ denoting the K{\"a}llen
function $\lambda(a,b,c)=(a-b-c)^2-4\,bc$. $s_{aij}$ is given in terms of
the evolution and splitting variables as $s_{aij}=y\,(q^2-m_k^2)+(1-y)\,(s_{ai}+m_j^2)$,
where
\begin{equation}
  y=\frac{t\,x_a/z_a}{q^2-s_{ai}-m_j^2-m_k^2}\;,\qquad
  \tilde{z}=\frac{z_a/x_a}{1-y}\,
  \frac{q^2-m_{aij}^2-m_k^2}{q^2-s_{ai}-m_j^2-m_k^2}\;.
\end{equation}
The new momentum of the emitter parton, $p_{ai}$, is constructed as
\begin{align}\label{eq:def_ff_pi_pj}
  p_{ai}^\mu\,=&\;\bar{z}_{ai}\,\frac{\gamma(q^2,s_{aij},m_k^2)\,p_{aij}^\mu
    -s_{aij}\,p_k^\mu}{\beta(q^2,s_{aij},m_k^2)}
  +\frac{s_{ai}+{\rm k}_\perp^2}{\bar{z}_{ai}}\,
  \frac{p_k^\mu-m_k^2/\gamma(q^2,s_{aij},m_k^2)\,p_{aij}^\mu}{
    \beta(q^2,s_{aij},m_k^2)}+k_\perp^\mu\;,
\end{align}
where $\beta(a,b,c)={\rm sgn}(a-b-c)\sqrt{\lambda(a,b,c)}$,
$2\,\gamma(a,b,c)=(a-b-c)+\beta(a,b,c)$ and $p_{aij}^\mu=q^\mu-p_k^\mu$.
The parameters $\bar{z}_{ai}$ and ${\rm k}_\perp^2=-k_\perp^2$ 
of this decomposition are given by
\begin{equation}\label{eq:def_ff_zi_kt}
  \begin{split}
    \bar{z}_{ai}\,=&\;\frac{q^2-s_{aij}-m_k^2}{\beta(q^2,s_{aij},m_k^2)}\,
    \left[\;\tilde{z}\,-\,\frac{m_k^2}{\gamma(q^2,s_{aij},m_k^2)}
      \frac{s_{aij}+s_{ai}-m_j^2}{q^2-s_{aij}-m_k^2}\right]\;,\\
        {\rm k}_\perp^2\,=&\;\bar{z}_{ai}\,(1-\bar{z}_{ai})\,s_{aij}-
        (1-\bar{z}_{ai})\, s_{ai}-\bar{z}_{ai}\, m_j^2\;,
  \end{split}
\end{equation}
The transverse momentum is constructed using an azimuthal angle, $\phi_{ai}$
\begin{equation}\label{eq:likt}
  k_\perp^\mu={\rm k}_\perp\left(\cos\phi_{ai} \frac{n_\perp^\mu}{|n_\perp|}
  +\sin\phi_{ai} \frac{l_\perp^{\,\mu}}{|l_\perp|}\right)\;,
  \qquad\text{where}\qquad
  n_\perp^\mu=\eps^{0\mu}_{\;\;\;\nu\rho}\,
  \tilde{p}_{aij}^{\,\nu}\,\tilde{p}_k^{\,\rho}\;,
  \qquad
  l_\perp^{\,\mu}=\eps^\mu_{\;\nu\rho\sigma}\,
  \tilde{p}_{aij}^{\,\nu}\,\tilde{p}_k^{\,\rho}\,n_\perp^\sigma\;.
\end{equation}
In kinematical configurations where $\vec{\tilde{p}}_{aij}=\pm\vec{\tilde{p}}_k$,
$n_\perp$ in the definition of Eq.~\eqref{eq:likt} vanishes. It can then be computed
as $n_\perp^\mu=\eps^{0\,i\mu}_{\;\;\;\;\;\nu}\,\tilde{p}_{aij}^{\,\nu}$,
where $i$ may be any index that yields a nonzero result.

The first branching step, which generates the final-state momentum $p_j$
and the intermediate momentum $p_{ai}$ is followed by a second step, constructed
using the same algorithm. As $p_{ai}$ has been generated with virtuality
$s_{ai}$, no momentum reshuffling is necessary in this case, and $p_k$ serves
as the defining vector for the anti-collinear direction only.
The customary variables $y$ and $\tilde{z}$ are determined by
\begin{equation}
  y'=\left[\,1+\frac{z_a}{x_a}\,\frac{
    q^2-m_{aij}^2-m_k^2}{
    s_{ai}-m_a^2-m_i^2}\right]^{-1}\;,\quad
  \tilde{z}'=x_a\;.
\end{equation}
Equations~\eqref{eq:def_ff_pi_pj} and~\eqref{eq:def_ff_zi_kt} are
employed to construct the momenta $p_a$ and $p_j$ using the replacements
$q^2\to (p_{ai}+p_k)^2$, $s_{aij}\to s_{ai}$, $s_{ai}\to m_a^2$ and
$m_j^2\to m_i^2$.

The phase-space factorization for final-state splittings with final-state
spectator can be derived similarly to the $2\to3$ case described
in~\cite{Dittmaier:1999mb}, App.~B. We perform an s-channel factorization
over $p_{aij}$ and subsequently over $p_{ai}$. This gives
\begin{equation}
  \begin{split}
    \int{\rm d}\Phi(p_a,p_i,p_j,p_k|\,q)
    =&\int\frac{{\rm d}s_{aij}}{2\pi}\,
    \int{\rm d}\Phi(p_{aij},p_k|\,q)\,
    \int{\rm d}\Phi(p_a,p_i,p_j|\,p_{aij})\\
    =&\int\frac{{\rm d}s_{aij}}{2\pi}\,
    \sqrt{\frac{\lambda(q^2,s_{aij},m_k^2)}{
        \lambda(q^2,m_{aij}^2,m_k^2)}}
    \int{\rm d}\Phi(\tilde{p}_{aij},\tilde{p}_k|\,q)\,
    \int{\rm d}\Phi(p_a,p_i,p_j|\,p_{aij})\\
    =&\int{\rm d}\Phi(\tilde{p}_{aij},\tilde{p}_k|\,q)\,
    \int\Big[{\rm d}\Phi(p_a,p_i,p_j|\,\tilde{p}_{aij},\tilde{p}_k)\Big]
  \end{split}    
\end{equation}
We define the auxiliary variable $\xi_a=z_a/x_a$ and use the relations
$z_a=(s_{ak}-m_a^2-m_k^2)/(q^2-m_{aij}^2-m_k^2)$,
$\xi_a=(s_{aik}-s_{ai}-m_k^2)/(q^2-m_{aij}^2-m_k^2)$
and $t=\xi_a\,(s_{aij}-s_{ai}-m_j^2)$ to write
\begin{equation}
  \begin{split}
    \int\Big[{\rm d}\Phi(p_a,p_i,p_j|\,\tilde{p}_{aij},\tilde{p}_k)\Big]
    =&\int\frac{{\rm d}s_{aij}}{2\pi}\,
    \sqrt{\frac{\lambda(q^2,s_{aij},m_k^2)}{
        \lambda(q^2,m_{aij}^2,m_k^2)}}
    \int\frac{{\rm d}s_{ai}}{2\pi}
    \int{\rm d}\Phi(p_{ai},p_j|p_{aij})
    \int{\rm d}\Phi(p_a,p_i|p_{ai})\\
    =&\int\frac{{\rm d}s_{aij}}{2\pi}\,
    \frac{1}{\sqrt{\lambda(q^2,m_{aij}^2,m_k^2)}}
    \int\frac{{\rm d}s_{ai}}{2\pi}
    \int\frac{{\rm d}s_{aik}\,{\rm d}\phi_{ai}}{4(2\pi)^2}
    \int\frac{1}{4(2\pi)^2}\frac{{\rm d}s_{ak}\,{\rm d}\phi_a}{
      \sqrt{\lambda(s_{aik},s_{ai},m_k^2)}}\\
    =&\frac{1}{4(2\pi)^3}
    \frac{q^2-m_{aij}^2-m_k^2}{\sqrt{\lambda(q^2,m_{aij}^2,m_k^2)}}
    \int{\rm d}t\int{\rm d}z_a\int{\rm d}\phi_{a}\\
    &\qquad\times\frac{1}{4(2\pi)^3}\int{\rm d}s_{ai}
    \int\frac{{\rm d}\xi_a}{\xi_a}\int{\rm d}\phi_{ai}\,
    \frac{q^2-m_{aij}^2-m_k^2}{\sqrt{\lambda(s_{aik},s_{ai},m_k^2)}}\;.
  \end{split}
\end{equation}
The final result is
\begin{equation}\label{eq:tc_ff_ps}
  \begin{split}
    \int\Big[{\rm d}\Phi(p_a,p_i,p_j|\,\tilde{p}_{aij},\tilde{p}_k)\Big]
    =&\,\frac{J^{(1)}_{\rm FF}}{16\pi^2}
    \int\frac{{\rm d}t}{t}\int{\rm d}z_a\int\frac{{\rm d}\phi_j}{2\pi}\,
    \bigg[\,\frac{1}{16\pi^2}
      \int{\rm d}s_{ai}\int\frac{{\rm d}\xi_a}{\xi_a}\int\frac{{\rm d}\phi_i}{2\pi}\,
      J^{(2)}_{\rm FF}\,\bigg]\,\frac{t}{\xi_a}\;,
  \end{split}  
\end{equation}
where we have defined the Jacobian factors
\begin{equation}
  J^{(1)}_{\rm FF}=\frac{q^2-m_{aij}^2-m_k^2}{
    \sqrt{\lambda(q^2,m_{aij}^2,m_k^2)}}
  \qquad\text{and}\qquad
  J^{(2)}_{\rm FF}=\frac{s_{aik}-s_{ai}-m_k^2}{
    \sqrt{\lambda(s_{aik},s_{ai},m_k^2)}}\;.
\end{equation}
The extension of Eq.~\eqref{eq:tc_ff_ps} to $D=4-2\eps$ dimensions
is straightforward. We obtain an additional factor of
\begin{equation}\label{eq:tc_ff_ps_dd}
  \tilde{\Delta}\Phi_{\rm FF}(p_a,p_i,p_j|\,\tilde{p}_{aij},\tilde{p}_k)=
  \left(\frac{\lambda(q^2,s_{aij},m_k^2)}{
    \lambda(q^2,m_{aij}^2,m_k^2)}\right)^{-\eps}
  \Delta\Phi_{\rm FF}(p_a,p_i,p_j|\,\tilde{p}_{aij},\tilde{p}_k)\;,
\end{equation}
where
\begin{equation}\label{eq:tc_ff_ps_dd_2}
  \Delta\Phi_{\rm FF}(p_a,p_i,p_j|\,\tilde{p}_{aij},\tilde{p}_k)=
  \left(\frac{\Omega(1-2\eps)}{(2\pi)^{-2\eps}}\right)^2
  \left(\bar{p}_{ai,j}^2\sin^2\theta_{ai,j}^{\,k}\sin^2\phi_j\right)^{-\eps}
  \left(\bar{p}_{a,i}^2\sin^2\theta_{a,i}^{\,k}\sin^2\phi_i\right)^{-\eps}\;.
\end{equation}
The $n$-dimensional sphere is defined as $\Omega(n)=2\pi^{n/2}/\,\Gamma(n/2)$.
We can write the magnitudes of the momenta as
\begin{equation}\label{eq:tc_ff_ps_dd_p}
  \bar{p}_{a,i}^2=\frac{\lambda(s_{ai},m_a^2,m_i^2)}{4\,s_{ai}}\;.
\end{equation}
The polar angles are given by
\begin{equation}\label{eq:tc_ff_ps_dd_costh}
  \cos\theta_{a,i}^{\,k}=
  -\frac{(s_{ai}+m_a^2-m_i^2)(s_{ai}+m_k^2-s_{aik})}{
    \sqrt{\lambda(s_{ai},m_a^2,m_i^2)\lambda(s_{ai},m_k^2,s_{aik})}}
  \left(1-\frac{2\,s_{ai}}{s_{ai}+m_a^2-m_i^2}\frac{p_ap_k}{p_{ai}p_k}\right)\;.
\end{equation}
The splitting functions in our algorithm are independent of $\phi_j$,
hence we can average over one azimuthal angle, leading to the familiar volume factor
\begin{equation}
  \frac{\Omega(2-2\eps)}{(2\pi)^{1-2\eps}}=\frac{\Omega(1-2\eps)}{(2\pi)^{-2\eps}}
  \int_0^\pi\frac{{\rm d}\phi_j}{2\pi}(\sin^2\phi_j)^{-\eps}=
  \frac{(4\,\pi)^{\eps}}{\Gamma(1-\eps)}\;.
\end{equation}
The azimuthal angle $\phi_i$ is parametrized as $\phi_i=\phi_{a,j}^{ai,k}$, where%
\footnote{Although we do not use this method in practice, it is instructive to show
  that we can use the technique of~\cite{Gehrmann-DeRidder:2003pne} to parametrize 
  the azimuthal angle integration by an auxiliary variable, $\chi$, defined as
  $ 
    s_{ij}=s_{ij,-}+\chi(s_{ij,+}-s_{ij,-})\;,
  $ 
  where $s_{ij,\pm}$ are the values of $s_{ij}$ at the phase-space boundaries,
  $\cos\phi_{a,j}^{ai,k}=\pm 1$. We obtain
  $ 
    \sin^2\phi_{a,j}^{ai,k}=
    4(s_{ij}-s_{ij,-})(s_{ij,+}-s_{ij})/(s_{ij,+}-s_{ij,-})^2
    =4\chi(1-\chi)\;.
  $ 
  The Jacobian factor related to this transformation is given by
  $ 
    {\rm d}\phi_{a,j}^{ai,k}/{\rm d}\chi=2\csc\phi_{a,j}^{ai,k}
    =(\chi(1-\chi))^{-1/2}\;.
  $ 
}
\begin{equation}\label{eq:tc_ff_ps_dd_cosphi}
  \cos\phi_{i,j}^{a,b}=\frac{\bar{s}_{ab}(\bar{s}_{ia}\bar{s}_{jb}+\bar{s}_{ib}\bar{s}_{ja}-\bar{s}_{ij}\bar{s}_{ab})
    -(s_a \bar{s}_{ib}\bar{s}_{jb}+s_b \bar{s}_{ia}\bar{s}_{ja}-\bar{s}_{ij}s_a s_b)}{
    \sqrt{k_\perp^2(p_i|\,p_a,p_b)\,k_\perp^2(p_j|\,p_a,p_b)}\;\lambda(s_{ab},s_a,s_b)/4}\;.
\end{equation}
Note that we have defined $\bar{s}_{ij}=p_ip_j$ to simplify the notation. The transverse momentum squared is given by
\begin{equation}
  k_\perp^2(p_i|p_a,p_b)=\frac{2\bar{s}_{ab}\bar{s}_{ia}\bar{s}_{ib}-s_a \bar{s}_{ib}^2-s_b \bar{s}_{ia}^2-s_i \bar{s}_{ab}^2
    +s_i s_a s_b}{\lambda(s_{ab},s_a,s_b)/4}\;.
\end{equation}
For massless partons, Eq.~\eqref{eq:tc_ff_ps_dd_costh} can be written in the simple form
\begin{equation}
  \cos\theta_{a,i}^k=1-2\,x_a\;,
  \qquad
  \cos\theta_{ai,j}^k=\frac{s_{aij}+s_{ai}}{s_{aij}-s_{ai}}
  \left[\,1-2\,\xi_a\,\frac{s_{aij}}{s_{aij}+s_{ai}}\,
    \frac{q^2}{q^2-s_{aij}}\right]\;.  
\end{equation}
The magnitudes of the momenta in this case are given by
\begin{equation}\label{eq:tc_ff_ps_pbar}
  4\bar{p}_{ai,j}^2=\frac{t}{\xi_a}\frac{t/\xi_a}{t/\xi_a+s_{ai}}\;,
  \qquad
  4\bar{p}_{a,i}^2=s_{ai}\;.
\end{equation}
In the iterated double collinear limit, we thus obtain the expected result
\begin{equation}\label{eq:dcdc_ps_fs}
  \Delta\Phi_{F}(s_{aij},s_{ai},z_a,x_a,\phi_i)=
  \frac{2\,(2\pi)^{2\eps}}{\Gamma(1-2\eps)}
  \Big(s_{aij}\,s_{ai}\,x_a(1-x_a)\,\xi_a(1-\xi_a)
  \sin^2\phi_{a,j}^{ai,k}\Big)^{-\eps}\;.
\end{equation}
This result is used in Sec.~\ref{sec:formalism}, Eq.~\eqref{eq:iterm_mc}
to derive the logarithmic contributions related to the phase-space integral.
It shows that our choice of variables correctly identifies $z_a$ and $x_a$
with light-cone momentum fractions in the collinear limit.

\subsection{Final-state emitter with initial-state spectator}
\label{sec:css_kin_fi_123}
\begin{figure}
  \includegraphics[scale=1]{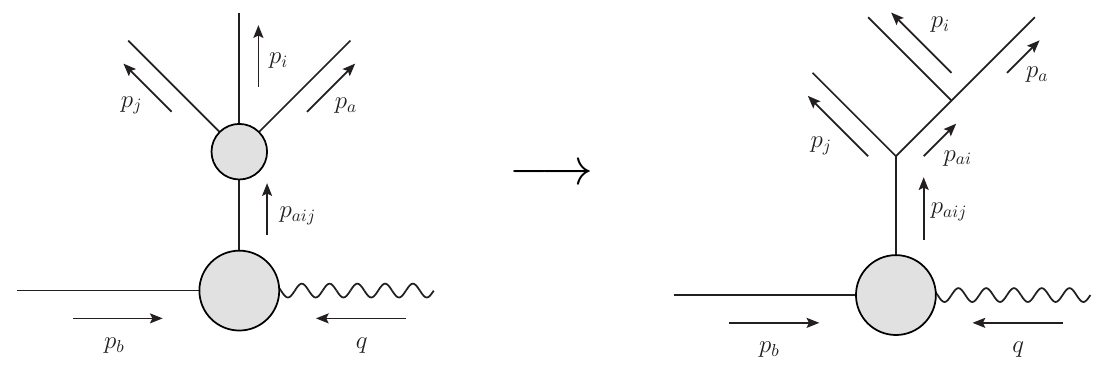}
  \caption{Kinematics mapping for final-state splittings with initial-state spectator.
    \label{fig:kin_fi_123}}
\end{figure}
Final-state splittings with initial-state spectator are treated in the same manner
as final-state splittings with final-state spectator, with the sole exception of
the construction of the new spectator momentum in the first branching step,
if the spectator is massive. The evolution and splitting variables are defined as
\begin{equation}\label{eq:def_var_if_123}
  t=\frac{2\,p_jp_{ai}\,p_{ai}p_b}{p_{aij}p_b}\;,\quad
  z_a=\frac{p_ap_b}{p_{aij}p_b}\;
  \qquad\text{and}\qquad
  s_{ai}\;,\quad
  x_a=\frac{p_ap_b}{p_{ai}p_b}\;
\end{equation}
The new spectator momentum is defined as
\begin{equation}\label{eq:def_fi_pa}
  \begin{split}
    p_b^{\,\mu}=&\;\left(\tilde{p}_b^{\,\mu}-\frac{q\cdot\tilde{p}_b}{q_\parallel^2}\,
    q_\parallel^{\,\mu}\right)\,
    \sqrt{\frac{\lambda(q^2,s_{aij},m_b^2)-4\,m_b^2\,\tilde{p}_{aij\perp}^2}{
        \lambda(q^2,m_{aij}^2,m_b^2)-4\,m_b^2\,\tilde{p}_{aij\perp}^2}}
    +\frac{q^2+m_b^2-s_{aij}}{2\,q_\parallel^2}\,q_\parallel^{\,\mu}\;,
  \end{split}
\end{equation}
where $q=\tilde{p}_b-\tilde{p}_{aij}$, $q_\parallel=q+\tilde{p}_{aij\perp}$ and
$s_{aij}=(x-1)/x\,(q^2-m_a^2)+(s_{ai}+m_j^2)/x$, where
\begin{equation}
  x=\left[\,1-\frac{t\,x_a/z_a}{q^2-s_{ai}-m_j^2-m_b^2}\right]^{-1}\;,\qquad
  \tilde{z}=\frac{z_a}{x_a}\;.
\end{equation}
The remaining construction proceeds as in Sec.~\ref{sec:css_kin_ff_123},
except that $m_k\to m_b$ and $p_k\to -p_b$. This is sketched
in Fig.~\ref{fig:kin_fi_123}. The customary variables $y$ and $\tilde{z}$
in the second branching step are given by
\begin{equation}
  y'=\left[\,1+\frac{
    t-z_a/x_a(q^2-s_{ai}-m_j^2-m_b^2)}{
    s_{ai}-m_a^2-m_i^2}\right]^{-1}\;,\quad
  \tilde{z}'=x_a\;.
\end{equation}

The phase-space factorization for final-state splittings with initial-state
spectator can be derived similarly to the $2\to3$ case described
in~\cite{Dittmaier:1999mb}, App.~B. We perform an s-channel factorization
over $p_{aij}$ and subsequently over $p_{ai}$. This gives
\begin{equation}
  \begin{split}
    \int{\rm d}\Phi(p_a,p_i,p_j,K|\,p_b,p_c)
    =&\int\frac{{\rm d}s_{aij}}{2\pi}\,
    \int{\rm d}\Phi(p_{aij},K|\,p_b,p_c)\,
    \int{\rm d}\Phi(p_a,p_i,p_j|\,p_{aij})\\
    =&\int{\rm d}\bar{x}
    \int{\rm d}\Phi(\tilde{p}_{aij},K|\,\tilde{p}_b,p_c)\,
    \int\Big[{\rm d}\Phi(p_a,p_i,p_j|\,p_b,p_c,q)\Big]
  \end{split}    
\end{equation}
where $\bar{x}=(q^2-m_{aij}^2-m_b^2)/(q^2-s_{aij}-m_b^2)$ and
\begin{equation}
  \begin{split}
    \int\Big[{\rm d}\Phi(p_a,p_i,p_j|\,p_b,p_c,q)\Big]
    =&\frac{\rho_{bai}}{2\pi}\,\frac{m_{aij}^2+m_b^2-q^2}{\bar{x}^2}
    \int\frac{{\rm d}s_{ai}}{2\pi}
    \int\frac{1}{4(2\pi)^2}\frac{{\rm d}s_{bai}\,{\rm d}\phi_{ai}}{
      \sqrt{\lambda(s_{aij},m_b^2,q^2)}}
    \int{\rm d}\Phi(p_a,p_i|\,p_{ai})\\
    =&\frac{1}{4(2\pi)^3}\,\frac{\rho_{bai}}{\bar{x}}\,
    \frac{s_{aij}+m_b^2-q^2}{\sqrt{\lambda(s_{aij},m_b^2,q^2)}}
    \int{\rm d}s_{bai}\int{\rm d}\phi_{ai}\,
    \int\Big[{\rm d}\Phi(p_a,p_i|\,p_{ai},p_b,q)\Big]\;.
  \end{split}  
\end{equation}
To simplify this expression, we have used the definition~\cite{Dittmaier:1999mb}
\begin{equation}
  \rho_{bai}=\sqrt{\frac{\lambda((\tilde{p}_b+p_c)^2,m_b^2,m_c^2)}{
      \lambda((p_b+p_c)^2,m_b^2,m_c^2)}}\;,
\end{equation}
We use the relation $z_a=(s_{ab}-m_a^2-m_b^2)/(q^2-s_{aij}-m_b^2)$ to write
\begin{equation}
  \begin{split}
    \int\Big[{\rm d}\Phi(p_a,p_i|\,p_{ai},p_b,q)\Big]
    =&\,\int\frac{{\rm d}s_{ai}}{2\pi}
    \int{\rm d}\Phi(p_a,p_i|\,p_{ai})
    =\frac{1}{4(2\pi)^3}\int{\rm d}s_{ai}
    \int{\rm d}z_a\int{\rm d}\phi_a\,
    \frac{s_{aij}+m_b^2-q^2}{\sqrt{\lambda(s_{ai},s_{bai},m_b^2)}}\;.
  \end{split}
\end{equation}
Using the auxiliary variable $\xi_a=z_a/x_a=(s_{bai}-s_{ai}-m_b^2)/(q^2-s_{aij}-m_b^2)$,
the final result can be written as
\begin{equation}\label{eq:tc_fi_ps}
  \begin{split}
    \int\Big[{\rm d}\Phi(p_a,p_i,p_j|\,p_b,p_c,q)\Big]
    =&\,\frac{J^{(1)}_{\rm FI}}{16\pi^2}
    \int{\rm d}z_a\int\frac{{\rm d}\phi_j}{2\pi}\,
    \bigg[\,\frac{1}{16\pi^2}\int{\rm d}s_{ai}
      \int\frac{{\rm d}\xi_a}{\xi_a}\int\frac{{\rm d}\phi_i}{2\pi}\,
      J^{(2)}_{\rm FI}\,\bigg]\,(s_{aij}+m_b^2-q^2)\;,
  \end{split}  
\end{equation}
where we have defined the Jacobian factors
\begin{equation}
  J^{(1)}_{\rm FI}=\frac{\rho_{bai}}{\bar{x}}\,
  \frac{s_{aij}+m_b^2-q^2}{
    \sqrt{\lambda(s_{aij},m_b^2,q^2)}}
  \qquad\text{and}\qquad
  J^{(2)}_{\rm FI}=\frac{s_{ai}+m_b^2-s_{bai}}{
    \sqrt{\lambda(s_{ai},m_b^2,s_{bai})}}\;.
\end{equation}
According to Eq.~\eqref{eq:def_var_if_123}, $s_{aij}$ (and therefore
$\bar{x}$ and $\rho_{bai}$) depends on both $t$ and $s_{ai}$, hence
$J^{(1)}_{\rm FI}$ is not independent of the second branching for
nonzero $m_b$. The evolution variable could be redefined as
$t=s_{aij}s_{aib}/(2p_{aij}p_b)$ to solve this problem. As we deal
with massless initial-state partons only, we defer this discussion
to a future publication.

The extension of Eq.~\eqref{eq:tc_fi_ps} to $D=4-2\eps$ dimensions
is straightforward. We obtain an additional factor of
\begin{equation}\label{eq:tc_fi_ps_dd}
  \Delta\Phi_{\rm FI}(p_a,p_i,p_j|\,\tilde{p}_b,\tilde{p}_c,q)=
  \left(\frac{\Omega(1-2\eps)}{(2\pi)^{-2\eps}}\right)^2
  \left(\bar{p}_{ai,j}^2\sin^2\theta_{ai,j}^{\,b}\sin^2\phi_j\right)^{-\eps}
  \left(\bar{p}_{a,i}^2\sin^2\theta_{a,i}^{\,b}\sin^2\phi_i\right)^{-\eps}\;.
\end{equation}
The momenta and polar angles are defined as in Eqs.~\eqref{eq:tc_ff_ps_dd_p}
and~\eqref{eq:tc_ff_ps_dd_costh}, and the azimuthal angle $\phi_i$ is parametrized
as $\phi_i=\phi_{a,j}^{ai,b}$, using Eq.~\eqref{eq:tc_ff_ps_dd_cosphi}.
As in the case of final-state emitter with final-state spectator, the splitting functions
are independent of $\phi_j$, hence we can average over one azimuthal angle.
For massless partons, the polar angles can be written in the simple form
\begin{equation}
  \cos\theta_{a,i}^b=1-2\,x_a\;,
  \qquad
  \cos\theta_{ai,j}^b=\frac{s_{aij}+s_{ai}}{s_{aij}-s_{ai}}
  \left[\,1-2\,\xi_a\,\frac{s_{aij}}{s_{aij}+s_{ai}}\right]\;.  
\end{equation}
The magnitudes of the momenta in this case are given by Eq.~\eqref{eq:tc_ff_ps_pbar}.
In the iterated collinear limit, Eq.~\eqref{eq:tc_fi_ps_dd} can be simplified
to give Eq.~\eqref{eq:dcdc_ps_fs}.

\subsection{Initial-state emitter with final-state spectator}
\label{sec:css_kin_if_123}
\begin{figure}
  \includegraphics[scale=1]{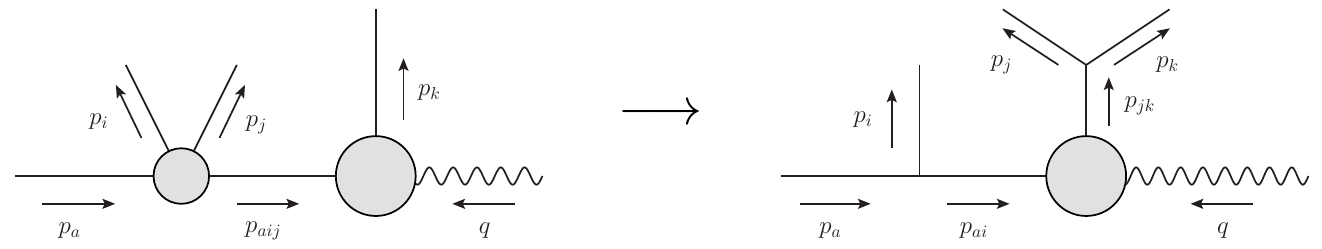}
  \caption{Kinematics mapping for initial-state splittings with final-state spectator.
    \label{fig:kin_if_123}}
\end{figure}
The kinematics in initial-state branchings with final-state spectator is typically
constructed by mapping the process to final-state branchings with initial-state
spectator~\cite{Catani:1996vz}. This mapping requires special care in the case of
$2\to4$ dipole splittings. Our algorithm is sketched in Fig.~\ref{fig:kin_if_123}.
We combine an initial-final branching\footnote{Both the global and the local recoil
  scheme, as defined in~\cite{Hoche:2015sya}, can be used. We describe only the global
  scheme in this publication.} during which the spectator is shifted off mass-shell
with a $1\to 2$ decay of the newly defined pseudoparticle with momentum $p_{jk}$.

We use the following evolution and splitting variables
\begin{equation}\label{eq:def_variables_if_123}
  \begin{split}
    t=\frac{2\,p_jp_{ai}\,p_{ai}p_k}{p_ap_{ijk}}\;,\quad
    z_a=\frac{-q^2}{2\,p_ap_{ijk}}
    \qquad\text{and}\qquad
    s_{ai}\;,\quad
    x_a=\frac{p_{ai}p_k}{p_ap_{ijk}}\;.
  \end{split}
\end{equation}
where $q=p_a-p_i-p_j-p_k$. We begin by constructing the initial-state branching.
As the spectator parton changes its virtuality, the shift in~\cite{Hoche:2015sya},
Eq.~(A.9) must be modified to
\begin{equation}\label{eq:def_if_pa_1}
  \begin{split}
    p_{jk}^{\,\mu}=&\;\left(\tilde{p}_k^{\,\mu}-\frac{q\cdot\tilde{p}_k}{q^2}\,
    q^{\,\mu}\right)\,
    \sqrt{\frac{\lambda(q^2,s_{ai},s_{jk})}{\lambda(q^2,m_{aij}^2,m_k^2)}}
    +\frac{q^2+s_{jk}-s_{ai}}{2\,q^2}\,q^{\,\mu}\;,
  \end{split}
\end{equation}
where $q=\tilde{p}_k-\tilde{p}_{aij}$ and $s_{jk}=q^2(1-x_a/z_a)+t/x_a-s_{ai}$.

Next we construct the momentum of the emitted particle, $p_i$, as
\begin{align}\label{eq:def_if_pi_pj}
  p_i^{\,\mu}\,=&\;-\bar{z}_i\,\frac{\gamma(q^2,s_{ai},s_{jk})\,
    p_{ai}^{\,\mu}+s_{ai}\,p_{jk}^{\,\mu}}{\beta(q^2,s_{ai},s_{jk})}+
  \frac{m_i^2+{\rm k}_\perp^2}{\bar{z}_i}\,\frac{
    p_{jk}^{\,\mu}+s_{jk}/\gamma(q^2,s_{ai},s_{jk})\,p_{ai}^{\,\mu}}{
    \beta(q^2,s_{ai},s_{jk})}+k_\perp^\mu\;,
\end{align}
The parameters $\bar{z}_i$ and ${\rm k}_\perp^2=-k_\perp^2$ 
of this decomposition are given by
\begin{equation}\label{eq:def_if_zi_kt}
  \begin{split}
    \bar{z}_i\,=&\;\frac{q^2-s_{ai}-s_{jk}}{\beta(q^2,s_{ai},s_{jk})}\,
    \left[\;\frac{x-1}{x-u}\,-\,
      \frac{s_{jk}}{\gamma(q^2,s_{ai},s_{jk})}
      \frac{s_{ai}+m_i^2-m_a^2}{q^2-s_{ai}-s_{jk}}\right]\;,\\
    {\rm k}_\perp^2\,=&\;\bar{z}_i\,(1-\bar{z}_i)\,s_{ai}
    -(1-\bar{z}_i)\,m_i^2-\bar{z}_i\, m_a^2\;,
  \end{split}
\end{equation}
where $u=-(s_{ai}-m_i^2-m_a^2)\,z_a/q^2$ and $x=u+x_a-t\,z_a/(q^2x_a)$.

We now boost $p_a$ and all final state particles into the frame
where $p_{a}$ is aligned along the beam direction, with $p_b$,
the opposite-side beam particle, unchanged. Eventually we must
construct the decay of the two-parton system defined by $p_{jk}$.
This can be achieved by using the same technique as in Sec.~\ref{sec:css_kin_ff_123},
i.e.\ we construct a decay with the customary variables $y$ and $\tilde{z}$
defined as
\begin{equation}
  y'=\left[1-\frac{t/x_a-q^2x_a/z_a}{s_{jk}-m_j^2-m_k^2}\right]^{-1}\;,
  \qquad
  \tilde{z}'=\frac{t/x_a}{t/x_a-q^2 x_a/z_a}\;.
\end{equation}
At the same time, we need to make the replacement $p_k\to -p_{ai}$,
$m_k\to s_{ai}$ and use the appropriate final-state masses.
This technique is sketched in Fig.~\ref{fig:kin_fi_123}.

The phase-space factorization for initial-state splittings with final-state
spectator can be derived similarly to the $2\to3$ case described
in~\cite{Dittmaier:1999mb}, App.~B. We first perform the s-channel
factorization over $p_{ijk}$. Using $z_a=-q^2/(s_{ijk}+m_a^2-q^2)$, this gives
\begin{equation}
  \begin{split}
    \int{\rm d}\Phi(p_i,p_j,p_k,K|\,p_a,p_b)
    =&\int\frac{{\rm d}s_{ijk}}{2\pi}\,
    \int{\rm d}\Phi(p_{ijk},K|\,p_a,p_b)\,
    \int{\rm d}\Phi(p_i,p_j,p_k|\,p_{ijk})\\
    =&\int{\rm d}z_a\,\int{\rm d}\Phi(\tilde{p}_k,K|\,\tilde{p}_a,p_b)\,
    \int\Big[{\rm d}\Phi(p_i,p_j,p_k|\,p_a,p_b,q)\Big]
  \end{split}    
\end{equation}
where
\begin{equation}
  \begin{split}
    \int\Big[{\rm d}\Phi(p_i,p_j,p_k|\,p_a,p_b,q)\Big]
    =&\frac{1}{2\pi}\,\frac{\rho_{ija}\,q^2}{z_a^2}
    \int\frac{{\rm d}s_{jk}}{2\pi}
    \int\frac{1}{4(2\pi)^2}\frac{{\rm d}s_{ai}\,{\rm d}\phi_i}{\sqrt{\lambda(s_{ijk},m_a^2,q^2)}}
    \int{\rm d}\Phi(p_j,p_k|p_{jk})\\
    =&\frac{1}{4(2\pi)^3}\,\frac{\rho_{ija}}{z_a}\,
    \frac{s_{ijk}-m_a^2-q^2}{\sqrt{\lambda(s_{ijk},m_a^2,q^2)}}
    \int{\rm d}s_{ai}\int{\rm d}\phi_i\,
    \int\Big[{\rm d}\Phi(p_j,p_k|p_{jk},p_a,q)\Big]\;.
  \end{split}  
\end{equation}
To simplify this expression, we have used the definition~\cite{Dittmaier:1999mb}
\begin{equation}
  \rho_{ija}=\sqrt{\frac{\lambda((\tilde{p}_{aij}+p_b)^2,m_a^2,m_b^2)}{
      \lambda((p_a+p_b)^2,m_a^2,m_b^2)}}\;,
\end{equation}
where $p_a^\mu$ is given by momentum conservation using Eq.~\eqref{eq:def_if_pa_1}.%
\footnote{Note that $p_a^\mu$ depends on the recoil scheme~\cite{Hoche:2015sya},
  and therefore $\rho_{ija}$ is generally scheme dependent. However, in the most
  relevant case of $m_a=m_{aij}=0$, i.e.\ for massless initial-state partons,
  we obtain $\rho_{ija}=z_a$.}
We make use of the relations $x_a=z_a\,(q^2-s_{aij}-s_{jk}+m_j^2)/q^2$
and $t=-x_a\,(s_{aij}-s_{ai}-m_j^2)$ to write
\begin{equation}
  \begin{split}
    \int\Big[{\rm d}\Phi(p_j,p_k|p_{jk},p_a,q)\Big]
    =&\,\int\frac{{\rm d}s_{jk}}{2\pi}
    \int{\rm d}\Phi(p_j,p_k|p_{jk})
    =\frac{1}{4(2\pi)^3}\int\frac{{\rm d}x_a}{x_a}
    \int{\rm d}t\int{\rm d}\phi_j\,
    \frac{-q^2/z_a}{\sqrt{\lambda(s_{jk},s_{ai},q^2)}}\;.
  \end{split}
\end{equation}
The final result is
\begin{equation}\label{eq:tc_if_ps}
  \begin{split}
    \int\Big[{\rm d}\Phi(p_i,p_j,p_k|\,p_a,p_b,q)\Big]
    =&\,\frac{J^{(1)}_{\rm IF}}{16\pi^2}
    \int\frac{{\rm d}t}{t}\int\frac{{\rm d}\phi_j}{2\pi}\,
    \bigg[\,\frac{1}{16\pi^2}\int{\rm d}s_{ai}
      \int\frac{{\rm d}x_a}{x_a}\int\frac{{\rm d}\phi_i}{2\pi}\,
      J^{(2)}_{\rm IF}\,\bigg]\,\frac{t}{x_a}\;,
  \end{split}  
\end{equation}
where we have defined the Jacobian factors
\begin{equation}
  J^{(1)}_{\rm IF}=\frac{\rho_{ija}}{z_a}\,
    \frac{s_{ijk}+m_a^2-q^2}{
    \sqrt{\lambda(s_{ijk},m_a^2,q^2)}}
  \qquad\text{and}\qquad
  J^{(2)}_{\rm IF}=
  \frac{-q^2\,x_a/z_a}{\sqrt{\lambda(s_{jk},s_{ai},q^2)}}\;.
\end{equation}
Note that $s_{ijk}=q^2(1-1/z_a)-m_a^2$, therefore both $\rho_{ija}$
and $J^{(1)}_{\rm IF}$ are unaffected by the intrinsic branching.

The extension of Eq.~\eqref{eq:tc_if_ps} to $D=4-2\eps$ dimensions
is straightforward. We obtain an additional factor of
\begin{equation}\label{eq:tc_if_ps_dd}
  \Delta\Phi_{\rm IF}(p_i,p_j,p_k|\,\tilde{p}_a,\tilde{p}_b,q)=
  \left(\frac{\Omega(1-2\eps)}{(2\pi)^{-2\eps}}\right)^2
  \left(\bar{p}_{j,k}^2\sin^2\theta_{j,k}^{\,ai}\sin^2\phi_j\right)^{-\eps}
  \left(\bar{p}_{i,jk}^2\sin^2\theta_{i,jk}^{\,a}\sin^2\phi_i\right)^{-\eps}\;.
\end{equation}
The momenta and polar angles are defined as in
Eqs.~\eqref{eq:tc_ff_ps_dd_p} and~\eqref{eq:tc_ff_ps_dd_costh},
and the azimuthal angle $\phi_i$ is parametrized as $\phi_i=\phi_{a,j}^{ai,jk}$,
using Eq.~\eqref{eq:tc_ff_ps_dd_cosphi}. 
As in the case of final-state emitter with final-state spectator, the splitting functions
are independent of $\phi_j$, hence we can average over one azimuthal angle.
For massless partons, the polar angles can be written in the simple form
\begin{equation}
  \cos\theta_{i,jk}^{\,a}=1-\frac{2\,z_a(1-z_a)s_{ai}}{q^2(1-x_a)+tz_a/x_a-z_as_{ai}}\;,
  \qquad
  \cos\theta_{j,k}^{\,ai}=\frac{1-2\,z_a t/(z_a t-q^2 x_a^2)}{
    \sqrt{1-4\,s_{ai}s_{jk}/(q^2x_a/z_a-t/x_a)^2}}\;.
\end{equation}
The magnitudes of the momenta in this case are given by
\begin{equation}
  4\bar{p}_{i,jk}^2=\frac{-q^2}{z_a(1-z_a)}
  \left[\,1-x_a+\frac{t/x_a-s_{ai}}{q^2/z_a}\right]^2\;,
  \qquad
  4\bar{p}_{j,k}^2=q^2\left(1-\frac{x_a}{z_a}\right)+\frac{t}{x_a}-s_{ai}\;.
\end{equation}
In the iterated double collinear limit, we thus obtain the expected result
\begin{equation}\label{eq:dcdc_ps_is}
  \Delta\Phi_{I}(s_{aij},s_{ai},z_a,x_a,\phi_i)=
  \frac{2\,(2\pi)^{2\eps}}{\Gamma(1-2\eps)}
  \Big(s_{aij}s_{ai}(1-x_a)(1-\xi_a)
  \sin^2\phi_{a,j}^{ai,jk}\Big)^{-\eps}\;.
\end{equation}
This result is used in Sec.~\ref{sec:formalism}, Eq.~\eqref{eq:iterm_mc}
to derive the logarithmic contributions related to the phase-space integral.
It shows that our choice of variables correctly identifies $z_a$ and $x_a$
with light-cone momentum fractions in the collinear limit.

\subsection{Initial-state emitter with initial-state spectator}
\label{sec:css_kin_ii_123}
\begin{figure}
  \includegraphics[scale=1]{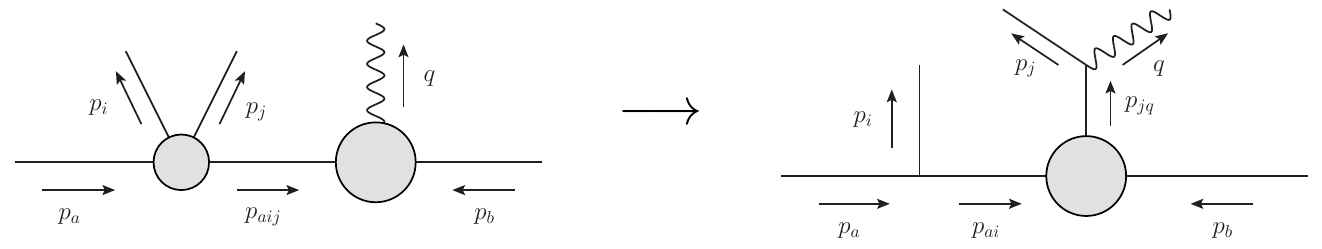}
  \caption{Kinematics mapping for initial-state splittings with initial-state spectator.
    \label{fig:kin_ii_123}}
\end{figure}
Similar to the case of initial-state splitters with final-state spectator,
the kinematics in initial-state branchings with initial-state spectator requires
special care in the case of $2\to4$ dipole splittings. Our algorithm is sketched
in Fig.~\ref{fig:kin_ii_123}. We combine an initial-initial branching during which
the final-state virtuality is promoted to $(p_j+q)^2$ with a $1\to 2$ decay of the
newly defined pseudoparticle into $p_j$ and $q$.

Our evolution and splitting variables are defined as
\begin{equation}\label{eq:def_variables_ii_123}
  \begin{split}
    t=\frac{2\,p_jp_{ai}\,p_{ai}p_b}{p_ap_b}\;,\quad
    z_a=\frac{q^2}{2\,p_ap_b}
    \qquad\text{and}\qquad
    s_{ai}\;,\quad
    x_a=\frac{p_{ai}p_b}{p_ap_b}\;.
  \end{split}
\end{equation}
We first determine the new momentum of the initial-state parton as
\begin{equation}\label{eq:def_ii_pk}
  \begin{split}
    p_a^{\,\mu}=&\;\left(\tilde{p}_{aij}^{\,\mu}-\frac{\tilde{m}_{aij}^2}{
      \gamma(q^2,\tilde{m}_{aij}^2,m_b^2)}\,p_b^{\,\mu}\right)\,
    \sqrt{\frac{\lambda(s_{ab},m_a^2,m_b^2)}{
        \lambda(q^2,\tilde{m}_{aij}^2,m_b^2)}}
    +\frac{m_a^2}{\gamma(s_{ab},m_a^2,m_b^2)}\,p_b^{\,\mu}\;,
  \end{split}
\end{equation}
where $q=p_a+p_b-p_i-p_j$ and $s_{ab}=q^2/z_a+m_a^2+m_b^2$.
Next we construct the momentum of the emitted parton, $p_j$, as
\begin{align}\label{eq:def_ii_pi_pj}
  p_i^{\,\mu}\,=&\;(1-\bar{z}_{ai})\,\frac{
    \gamma(s_{ab},m_a^2,m_b^2)\,p_a^{\,\mu}-m_a^2\,p_b^{\,\mu}}{
    \beta(s_{ab},m_a^2,m_b^2)}+
  \frac{m_i^2+{\rm k}_\perp^2}{1-\bar{z}_{ai}}\,\frac{
    p_b^{\,\mu}-m_b^2/\gamma(s_{ab},m_a^2,m_b^2)\,p_a^{\,\mu}}{
    \beta(s_{ab},m_a^2,m_b^2)}-k_\perp^\mu\;,
\end{align}
The parameters $\bar{z}_{aij}$ and ${\rm k}_\perp^2=-k_\perp^2$ 
of this decomposition are given by
\begin{equation}\label{eq:def_ii_zi_kt}
  \begin{split}
    \bar{z}_{ai}\,=&\;\frac{s_{ab}-m_a^2-m_b^2}{\beta(s_{ab},m_a^2,m_b^2)}\,
    \left[\;x_a\,-\,\frac{m_b^2}{\gamma(s_{ab},m_a^2,m_b^2)}
      \frac{s_{ai}+m_a^2-m_i^2}{s_{ab}-m_a^2-m_b^2}\right]\;,\\
    {\rm k}_\perp^2\,=&\;\bar{z}_{ai}\,(1-\bar{z}_{ai})\,m_a^2
    -(1-\bar{z}_{ai})\,s_{ai}-\bar{z}_{ai}\,m_i^2\;,
  \end{split}
\end{equation}
In a second step, we branch the new final state momentum $p_{ai}+p_b$ into
$p_j$ and $q$, using the spectator $p_{ai}$ and satisfying the constraint
$q^2=\tilde{q}^2$, where $\tilde{q}=\tilde{p}_{aij}+p_b$.
We employ the kinematics mapping of Sec.~\ref{sec:css_kin_ff_123}.
The customary variables $y$ and $\tilde{z}$ in this case are defined as
\begin{equation}
  y'=\left[1+\frac{q^2\,x_a/z_a+2s_{ai}}{
      q^2(x_a/z_a-1)+s_{ai}+m_b^2-m_j^2}\right]^{-1}\;,\qquad
  \tilde{z}'=\frac{t/x_a}{q^2\,x_a/z_a+2s_{ai}}\;.
\end{equation}
At the same time, we make the replacement $p_k\to p_{ai}$,
$m_k\to s_{ai}$ and use the appropriate final-state masses.
Finally we boost all remaining final-state particles into the frame defined by $q$,
using the algorithm defined in Sec.~(5.5) of~\cite{Catani:1996vz}.
The Lorentz transformation, $\Lambda$, is computed as
\begin{equation}\label{eq:def_ii_boost_123}
  \Lambda(\tilde{q},q)^\mu_{\;\nu}=g^\mu_{\;\nu}
  -\frac{2\,(q+\tilde{q})^\mu(q+\tilde{q})_\nu}{(q+\tilde{q})^2}
  +\frac{2\,q^\mu\tilde{q}_\nu}{\tilde{q}^2}\;,
\end{equation}

The phase-space factorization for initial-state splittings with
initial-state spectator can be derived similar to the $2\to3$
case described in~\cite{Dittmaier:1999mb}, App.~B.
We first perform the s-channel factorization over $p_{ijk}$.
Using $z_a=q^2/(s_{ijq}-m_a^2-m_b^2)$, this gives
\begin{equation}
  \begin{split}
    \int{\rm d}\Phi(p_i,p_j,q|\,p_a,p_b)
    =&\int\frac{{\rm d}s_{ijq}}{2\pi}\,
    \int{\rm d}\Phi(p_{ijq}|\,p_a,p_b)\,
    \int{\rm d}\Phi(p_i,p_j,q|\,p_{ijq})\\
    =&\int{\rm d}z_a\,\int{\rm d}\Phi(\tilde{q}|\,\tilde{p}_a,p_b)\,
    \int\Big[{\rm d}\Phi(p_i,p_j,q|\,p_a,p_b)\Big]
  \end{split}    
\end{equation}
where
\begin{equation}
  \begin{split}
    \int\Big[{\rm d}\Phi(p_i,p_j,q|\,p_a,p_b)\Big]
    =&\frac{1}{2\pi}\frac{q^2}{z_a}\,\int\frac{{\rm d}s_{jq}}{2\pi}
    \int\frac{1}{4(2\pi)^2}\frac{{\rm d}s_{ai}\,{\rm d}\phi_i}{
      \sqrt{\lambda(s_{ab},m_a^2,m_b^2)}}
    \int{\rm d}\Phi(p_j,q|p_{jq})\\
    =&\frac{1}{4(2\pi)^3}
    \frac{q^2/z_a}{\sqrt{\lambda(s_{ab},m_a^2,m_b^2)}}
    \int{\rm d}s_{ai}\int{\rm d}\phi_i\,
    \int\Big[{\rm d}\Phi(p_j,q|p_{jq},p_a,p_b)\Big]\;.
  \end{split}  
\end{equation}
We make use of the relations $x_a=z_a\,(s_{jq}-s_{ai}-m_b^2)/q^2$
and $t=-x_a\,(s_{aij}-s_{ai}-m_j^2)$ to write
\begin{equation}
  \begin{split}
    \int\Big[{\rm d}\Phi(p_j,q|p_{jq},p_a,p_b)\Big]
    =&\,\int\frac{{\rm d}s_{jq}}{2\pi}
    \int{\rm d}\Phi(p_j,q|p_{jq})
    =\frac{1}{4(2\pi)^3}\int\frac{{\rm d}x_a}{x_a}
    \int{\rm d}t\int{\rm d}\phi_j\,
    \frac{q^2/z_a}{\sqrt{\lambda(s_{jq},s_{ai},m_b^2)}}\;.
  \end{split}
\end{equation}
The final result is
\begin{equation}\label{eq:tc_ii_ps}
  \begin{split}
    \int\Big[{\rm d}\Phi(p_i,p_j,q|\,p_a,p_b)\Big]
    =&\,\frac{J^{(1)}_{\rm II}}{4(2\pi)^3}
    \int\frac{{\rm d}t}{t}\int{\rm d}\phi_j\,
    \bigg[\,\frac{1}{4(2\pi)^3}
      \int{\rm d}s_{ai}\int\frac{{\rm d}x_a}{x_a}\int{\rm d}\phi_i\,
      J^{(2)}_{\rm II}\,\bigg]\,\frac{t}{x_a}\;,
  \end{split}  
\end{equation}
where we have defined the Jacobian factors
\begin{equation}
  J^{(1)}_{\rm II}=\frac{s_{ab}-m_a^2-m_b^2}{
    \sqrt{\lambda(s_{ab},m_a^2,m_b^2)}}
  \qquad\text{and}\qquad
  J^{(2)}_{\rm II}=\frac{s_{jq}-s_{ai}-m_b^2}{
    \sqrt{\lambda(s_{jq},s_{ai},m_b^2)}}\;,
\end{equation}
and where $s_{jq}=q^2x_a/z_a+s_{ai}+m_b^2$.

The extension of Eq.~\eqref{eq:tc_ii_ps} to $D=4-2\eps$ dimensions
is straightforward. We obtain an additional factor of
\begin{equation}\label{eq:tc_ii_ps_dd}
  \Delta\Phi_{\rm II}(p_i,p_j,q|\,\tilde{p}_a,\tilde{p}_b)=
  \left(\frac{\Omega(1-2\eps)}{(2\pi)^{-2\eps}}\right)^2
  \left(\bar{p}_{j,q}^2\sin^2\theta_{j,q}^{\,ai}\sin^2\phi_j\right)^{-\eps}
  \left(\bar{p}_{i,jq}^2\sin^2\theta_{i,jq}^{\,a}\sin^2\phi_i\right)^{-\eps}\;.
\end{equation}
The momenta and polar angles are defined as in
Eqs.~\eqref{eq:tc_ff_ps_dd_p} and~\eqref{eq:tc_ff_ps_dd_costh},
and the azimuthal angle $\phi_i$ is parametrized as $\phi_i=\phi_{a,j}^{ai,jq}$,
using Eq.~\eqref{eq:tc_ff_ps_dd_cosphi}. 
As in the case of final-state emitter with final-state spectator, the splitting functions
are independent of $\phi_j$, hence we can average over one azimuthal angle.
For massless partons, the polar angles can be written in the simple form
\begin{equation}
  \cos\theta_{i,jq}^{\,a}=1+\frac{2\,s_{ai}}{(1-x_a)\,q^2/z_a-s_{ai}}\;,
  \qquad
  \cos\theta_{j,q}^{\,ai}=\frac{s_{jq}+s_{ai}}{s_{jq}-s_{ai}}
  \left(1-2\,\frac{s_{jq}}{s_{jq}-q^2}\,\frac{t/x_a}{q^2 x_a/z_a+2s_{ai}}\right)\,.
\end{equation}
The magnitudes of the momenta in this case are given by
\begin{equation}
  4\bar{p}_{i,jq}^2=\frac{q^2}{z_a}\left[\,1-x_a-z_a\frac{s_{ai}}{q^2}\right]^2\;,
  \qquad
  4\bar{p}_{j,q}^2=\left(q^2\frac{x_a}{z_a}+s_{ai}\right)\left[1-\frac{1}{x_a/z_a+s_{ai}/q^2}\right]^2\,.
\end{equation}
In the iterated collinear limit, Eq.~\eqref{eq:tc_ii_ps_dd} can be simplified
to give Eq.~\eqref{eq:dcdc_ps_is}.

\bibliography{journal}
\end{document}